\documentclass[11pt]{article}
\usepackage{graphicx}
\usepackage{amsmath}
\usepackage{amsxtra}
\newcommand{\word}[1]{\,\,\mbox{#1}\,\,}
\newcommand{\reff}[1]{(\ref{#1})}
\newcommand{\beq}{\begin{equation}}
\newcommand{\eeq}[1]{\label{#1}\end{equation}}
\newcommand{\beg}{\begin{equation*}}
\newcommand{\eeg}{\end{equation*}}
\newcommand{\negb}{\,\mid\!\!\!\not \;\;\;}

\newcommand{\expo}[1]{\mbox{e}^{#1}}

\newcommand{\sumprime}{\sideset{}{'}\sum}
\newcommand{\sumjq}{\sideset{}{^{j-q}}\sum}

\newcommand{\sumdq}{\sideset{}{^{d-q-1}}\sum}
\newcommand{\sumtwo}{\sideset{}{^2}\sum}
\newcommand{\bsplit}{\begin{split}}
\newcommand{\esplit}{\end{split}}

\newcommand{\bino}[2]{\Bigl(\genfrac{}{}{0pt}{}{#1}{#2}\Bigr)}
\begin{document}
\title{Multidimensional cut-off technique, odd-dimensional Epstein zeta functions and Casimir energy of massless scalar fields}
\author{\thanks{Email: aedery@ubishops.ca} Ariel Edery\\Bishop's University, Physics
Department\\Lennoxville, Quebec\\J1M1Z7}
\date{}
\maketitle
\begin{abstract}
\noindent Quantum fluctuations of massless scalar fields represented
by quantum fluctuations of the quasiparticle vacuum in a
zero-temperature dilute Bose-Einstein condensate may well provide
the first experimental arena for measuring the Casimir force of a
field other than the electromagnetic field. This would constitute a
real Casimir force measurement - due to quantum fluctuations - in
contrast to thermal fluctuation effects. We develop a
multidimensional cut-off technique for calculating the Casimir
energy of massless scalar fields in $d$-dimensional rectangular
spaces with $q$ large dimensions and $d-q$ dimensions of length $L$
and generalize the technique to arbitrary lengths. We explicitly
evaluate the multidimensional remainder and express it in a form
that converges exponentially fast. Together with the compact
analytical formulas we derive, the numerical results are exact and
easy to obtain. Most importantly, we show that the division between
analytical and remainder is not arbitrary but has a natural physical
interpretation. The analytical part can be viewed as the sum of
individual parallel plate energies and the remainder as an
interaction energy. In a separate procedure, via results from number
theory, we express some odd-dimensional homogeneous Epstein zeta
functions as products of one-dimensional sums plus a tiny remainder
and calculate from them the Casimir energy via zeta function
regularization.

\end{abstract}

\newpage

\section{Introduction}

The Casimir force remained for a long time one of the more esoteric
forces in Physics attracting at best some theoretical interest. All
of this has changed in the last eight years or so. After nearly $50$
years since its prediction in 1948 by Casimir \cite{Casimir}, the
force has now been successfully measured by a modern series of
experiments starting with Lamoreaux's 1997 landmark experiment
\cite{Lamoreaux} with a torsion pendulum which reduced errors
dramatically compared to the early 1958 experiment by Spaarnay
\cite{Spaarnay}. The force was subsequently measured more precisely
in 1998 using an atomic force microscope \cite{Mohideen} and the
measurements agreed with theoretical predictions to within 1\% after
finite conductivity, roughness and temperature corrections were
taken into account. Thus the modern era of precise Casimir
measurements was born and a non-exhaustive list of other
experimental studies since then can be found in
\cite{Klim,Roy1,Roy2,Harris,Chen1,Ederth,Bressi,Chen2,Decca}.
Interest in the Casimir force has also been fueled by theories with
large extra dimensions which predict among other things a deviation
from Newtonian gravitation at the sub-millimeter scale \cite{Dvali}.
To date no deviation has been found. Recently, a Casimir force
experiment \cite{Lopez} has placed new constraints on the parameters
of such proposed theories. An up-to-date list of gravitational
experiments can also be found in \cite{Lopez}. As with many
fundamental Physics discoveries, at first the Casimir force seemed
to have no apparent engineering application (since it is significant
only on micron or submicron scales). However, our ever increasing
ability to build structures on smaller scales has made the Casimir
force something various industries need to take into account. For
example, in 2001, scientists at Lucent Technologies showed that the
Casimir force could be used to control the mechanical motion of a
microelectromechanical system (MEMS) device \cite{Chan}(see also the
recent paper \cite{Janina} and references therein). MEMS are
micron-sized devices in which tiny sensors and actuators are carved
into a silicon substrate and are currently in use as car air-bag
sensors. For more details on the Casimir effect the reader is
referred to the following books \cite{Kimball,Mostepanenko} and
reviews \cite{Kimball2,Bordag,Barton,Jaeckel}.

All the measurements of the Casimir force to date have been limited
to the case of the electromagnetic field. However, experiments may
soon (or may already have done so indirectly) measure the Casimir
force for a massless scalar field. Quantum fluctuations of the
quasiparticle vacuum in a zero-temperature dilute Bose-Einstein
condensate (BEC) should give rise to a measurable Casimir force as
explained in recent papers \cite{Roberts2,Roberts}. The authors in
\cite{Roberts2,Roberts} state that indirect effects from these
quantum fluctuations may have already been observed
\cite{Pitaevskii,Stampur,Greiner,Vogels}. Note that this is a real
Casimir effect due to quantum fluctuations in contrast to thermal
fluctuations (often called pseudo-Casimir). The fact that the field
propagates at the speed of sound in the BEC medium in contrast to
the speed of light in Minkowski spacetime does not change anything
fundamental in relation to the Casimir energy. If the speed of
propagation is constant in a given medium, the Casimir energy in
units of this speed will be the same value regardless of whether the
medium is spacetime or a BEC. Moreover, a generally covariant action
analogous to what we see in General Relativity exists for scalar
fields propagating in a particular fluid. The Lagrangian is similar
to that of a massless Klein-Gordon field with the Minkowski metric
$\eta_{\mu\,\nu}$ of spacetime replaced by an effective or acoustic
metric $g_{\mu\,\nu}$ \cite{Visser}. Quoting directly from
\cite{Visser2}, ``at low momenta linearized excitations of the phase
of the condensate wavefunction obey a (3+1)-dimensional
d'Alembertian equation coupling to a (3+1)-dimensional
Lorentzian-signature `effective metric' that is generic, and depends
algebraically on the background field.". In \cite{Roberts} the
authors make the important observation that though the dispersion
relation for quantum fluctuations in a BEC is nonlinear, the Casimir
energy picks out mostly the long wavelength linear behaviour. This
is why the Casimir force $F_{BEC}$ calculated by the same authors
\cite{Roberts2,Roberts} for infinitely thin and infinitely repulsive
plates immersed in a zero-temperature three-dimensional dilute
condensate turns out to leading order to be the same as that of a
massless scalar field moving with the speed of sound $v$.

In this paper we are interested in the Casimir effect of massless
scalar fields traveling with speed $v$ in rectangular cavities of
$d$ spatial dimensions where $q$ dimensions are large and $d-q$
dimensions are of equal length $L$. The case of arbitrary lengths is
also considered in appendix B. We develop a multidimensional cut-off
technique to solve this problem. Why use a cut-off technique?
Clearly, it is less efficient than the zeta function technique that
yields quickly, via analytic continuation, finite results for
rectangular cavities in terms of Epstein zeta functions. There are a
few reasons for the importance of the exponential cut-off technique.
First, it remains the most physically intuitive method. For this
reason, recent texts in String Theory or Quantum Field Theory (QFT)
as well as courses in QFT introduce the standard Casimir energy
calculation of a string or parallel plates using an exponential
cut-off. For example, in the text {\it String Theory, Vol. I }
\cite{Polchinski}, the Casimir energy for the Bosonic string is
handled with an exponential cut-off. The result $\sum_{n=1}^{\infty}
n \to \tfrac{-1}{12}$ is obtained by replacing $n$ by
$n\,\expo{-\lambda\,n}$ and extracting the finite result
$\tfrac{-1}{12}$ from the series $\tfrac{1}{\lambda^2}
-\tfrac{1}{12} + O (\lambda^2)$. This cut-off method was used
instead of the zeta function technique which yields quickly
$\zeta(-1) =\tfrac{-1}{12}$. In his recent book, {\it Quantum Field
Theory in a Nutshell }\cite{Zee}, Zee brings in some humour in
explaining a Physicist's perspective on the same sum. I quote from
p.$66$, ``Aagh! What do we do with $\sum_{n=1}^{\infty} n$? None of
the ancient Greeks from Zeno on could tell us. What they should tell
us is that we are doing Physics...Physical plates cannot keep
arbitrarily high frequencies from leaking out.". He then introduces
the exponential cut-off to damp the ultraviolet frequencies. In the
classic QFT text by Itzykson and Zuber \cite{Zuber} the
electromagnetic parallel plate problem in three dimensions is solved
via a cut-off function and the Euler-Maclaurin formula and the same
technique can be seen applied in recent graduate courses (e.g. see
``Relativistic Quantum Field Theory I, Spring $2003$"
\cite{GuthMIT}). Physicists are therefore likely to be familiar with
the cut-off technique. Secondly, a multidimensional cut-off
calculation with an exact determination of the multidimensional
remainder term does not seem to have been systematically carried out
for rectangular cavities in arbitrary $d$ dimensions. Papers on
Casimir energies in arbitrary $d$ dimensions in rectangular cavities
have made use of dimensional and zeta function regularization
\cite{Wolfram,Neto,Li}. Explicit formulas using the exponential
cut-off technique in rectangular cavities include parallel plates in
higher dimensions \cite{Svaiter2}, rectangular cavities in two and
three dimensions \cite{Svaiter3,Lukosz,Trunov1,Trunov2}, and
explicit formulas via Poisson's formula up to $d=2$ appear in
\cite{Beneventano}. In \cite{Svaiter2,Svaiter3,Beneventano} the
connection between cut-off and zeta function technique is also
elaborated and explained. A detailed numerical analysis for the
electromagnetic case in three-dimensional rectangular cavities can
be found in \cite{Maclay}. Last but not least, by applying the
cut-off technique to rectangular cavities we are led in a natural
fashion to excellent finite analytical formulas plus a remainder. We
show that the division between analytical and remainder is not some
ad-hoc division. The analytical part has a clear physical
interpretation as sums of parallel plates out of which the
rectangular cavity is constructed. Moreover, the numerical results
are excellent because the analytical part is trivial to evaluate and
the multidimensional remainder is derived in a form that converges
quickly (exponentially fast). As already mentioned, the zeta
function technique applied to rectangular spaces has the great
advantage of leading quickly to finite results expressed in terms of
Epstein zeta functions. However, one then needs to go a few steps
further if one wants to express these in a convenient analytical
form and this is usually a separate procedure. In contrast,
analytical results are often a natural spin-off of the cut-off
technique.

One section of this paper is devoted to developing a technique that
derives highly accurate analytical formulas for a few
odd-dimensional homogeneous Epstein zeta functions. It turns out
that in even dimensions less than or equal to $8$ one can obtain
compact analytical expressions for the homogeneous Epstein zeta
function purely in terms of products of one-dimensional sums. There
is no remainder for these cases. This can be accomplished via number
theoretic formulas for the representation of integers as a sum of
squares in even dimensions. For even dimensions above $8$, the
number theoretic formulas get more complicated and in odd dimensions
above $7$ they are not presently known. For $3$, $5$ and $7$
dimensions the number theoretic formulas have only recently been
found \cite{Goro} but they are much more complicated than in even
dimensions. We therefore develop a procedure that uses the exact
even-dimensional results from number theory and then apply the
Euler-Maclaurin formula to obtain the odd dimensions. This yields
the homogeneous Epstein zeta function in $3$, $5$ and $7$ dimensions
as a finite number of products of one-dimensional sums plus a small
remainder term. This remainder is even smaller than the remainder
obtained via our multidimensional cut-off technique. For the most
important case of $3$ dimensions, we obtain both a highly compact
and extremely accurate analytical expression that contains only four
terms and where the remainder is a negligible $0.04\%$ of the
Casimir energy. Our specific procedure leads to low remainders but
is limited to a few homogeneous Epstein zeta functions, albeit one
that includes the three-dimensional case. A different more general
procedure applicable to any multidimensional inhomogeneous
Epstein-type zeta function can be found in \cite{Elizalde2}.

\section{Multidimensional cut-off technique including
remainder}

In this section we develop a multidimensional cut-off technique to
obtain formulas for the Casimir energy of a massless scalar field
$\phi(x)$ moving with a wave velocity $v$ in a $d$-dimensional
rectangular cavity with $d-q$ sides of equal length $L$ and $q$
sides of much larger length $L_m>>L$ where $m$ runs from $1$ to $q$.
One can generalize our method to arbitrary lengths and this is done
in appendix B. Here and throughout the paper we consider the more
special case as it makes the method, the formulas and the physical
interpretation more transparent. This section and appendix A (where
the remainder is evaluated) go together.

We consider periodic, Neumann and Dirichlet boundary conditions. The
fields are assumed to propagate in a homogeneous medium with a
constant speed $v$ and with a wavelength long enough that the
dispersion relation is linear i.e. $\omega =v\,k$ where $k$ is the
wavenumber. In other words, we assume the scalar field $\phi(x)$ to
obey the standard linear wave equation: \beq
\dfrac{\partial^{2}\phi(x)}{\partial^{2}\,t} - v^{2}
\nabla^{2}\phi(x)= 0. \eeq{sound} The boundary conditions are either
periodic, $\phi(x^i\!=\!0) =\phi(x^i\!=\!L)$, Neumann,
$\partial^i\phi(x)=0$ at $x^i\!=\!0$ and $x^i\!=\!L$ or Dirichlet
$\phi(x^i\!=\!0)=\phi(x^i\!=\!L)=0$. Here $i$ runs from $1$ to $d-q$
inclusively. After the standard fourier decomposition one obtains
the following quantized frequencies $\omega$ for periodic (p),
Neumann (N) and Dirichlet (D) conditions :
 \beq
\begin{split}
&\omega_p = 2\pi\,v \,(\tfrac{n_1^2}{L^2} + \cdots +
\tfrac{n_{d-q}^2}{L^2} + \tfrac{n_{d-q+1}^2}{L_1^2} + \cdots +
\tfrac{n_{d}^2}{L_q^2})^{1/2} \\& \omega_{N,D} = \pi\,v
(\tfrac{n_1^2}{L^2} + \cdots +
\tfrac{n_{d-q}^2}{L^2} + \tfrac{n_{d-q+1}^2}{L_1^2} + \cdots +
\tfrac{n_{d}^2}{L_q^2})^{1/2}
\end{split}
\eeq{omega} where the $n_i$'s run from $-\infty$ to $\infty$ for
periodic boundary conditions, $0$ to $\infty$ for Neumann and $1$ to
$\infty$ for Dirichlet. From quantum field theory we know that after
quantization the vacuum energy is given by the sum over all modes of
$\tfrac{1}{2}\,\omega$ (we work in units where $\hbar=1$). The
vacuum energies $E^{\,vac}$ for the three boundary conditions
labeled (p,N,D) are therefore: \beq
\begin{split}
&E^{\,vac}_{p}= \dfrac{\pi \,v}{L}\sum_{\substack{n_i=-\infty\\i=1,\ldots, d}}^{\infty} (n_1^2 + \cdots +
n_{d-q}^2 + \tfrac{n_{d-q+1}^2 L^2}{L_1^2} + \cdots +
\tfrac{n_{d}^2\,L^2}{L_q^2})^{1/2}\\&
 E^{\,vac}_{N,D}= \dfrac{\pi\,v}{2\,L} \,\sum_{\substack{n_i=0,1\\i=1,\ldots, d}}^{\infty} (n_1^2 + \cdots +
n_{d-q}^2 + \tfrac{n_{d-q+1}^2 L^2}{L_1^2} + \cdots +
\tfrac{n_{d}^2 L^2}{L_q^2})^{1/2}
\end{split}
\eeq{vacuum} The above sums are ultraviolet divergent and require
regularization. There are many different regularization schemes such
as exponential cut-off, zeta function and dimensional
regularization. In this paper the goal is to develop a
multidimensional cut-off technique via the Euler-Maclaurin formula.
Via this technique, we obtain formulas for the Casimir energy as a
finite sum over analytical terms plus a remainder. We fully evaluate
the remainder term and express it as sums over Bessel functions. We
later show that the analytical part has an intuitive physical
picture: it is the energy needed to construct the rectangular cavity
out of adding successive parallel plates.  We begin by calculating
the regularized vacuum energy for periodic boundary conditions.
After regularization, we then extract the finite Casimir energy
$E_p$ which is the difference between the regularized energy with
boundaries (discrete modes) minus the regularized energy without
boundaries (continuous modes). We later compare $E_p$ to the Epstein
zeta function obtained via zeta function regularization. We can
express the Neumann and Dirichlet energies, $E_N$ and  $E_D$, in
terms of sums over $E_p$ so only the periodic case needs to be
evaluated fully. The regularized vacuum energy $E^{\,reg}$ for
periodic boundary conditions using an exponential cut-off is: \beq
\begin{split}
&E^{\,reg}_p(q,\lambda) = \dfrac{\pi\,v}{L}\sum_{\substack{n_i=-\infty\\i=1,\ldots, d}}^{\infty} (n_1^2 + \cdots +
n_{d-q}^2 + \tfrac{n_{d-q+1}^2 L^2}{L_1^2} + \cdots +
\tfrac{n_{d}^2 L^2}{L_q^2})^{1/2}\\&\qquad\qquad\qquad\qquad\qquad\qquad\qquad\expo{-\lambda\,
\sqrt{n_1^2 + \cdots + n_{d-q}^2 + \frac{n_{d-q+1}^2 L^2}{L_1^2} +
\cdots + \frac{n_{d}^2 L^2}{L_q^2}}} \\
&= -\dfrac{\pi\,v}{L^{q+1}}\,\prod_{i=1}^q L_i\,\,\,\partial_{\lambda}\!\!\!\!\!
\sum_{\substack{n_i=-\infty\\i=1,\ldots, d-q}}^{\infty} \int_{-\infty}^{\infty}\expo{-\lambda\,\sqrt{n_1^2 + \cdots +
n_{d-q}^2 + x_1^2 + \cdots + x_q^2}}\,dx_1 \ldots dx_q
\end{split}
\eeq{start2} where we replaced the sums over the $q$ large
dimensions by integration. The parameter $\lambda$ is a free
parameter which we later set to $0$. The goal is to evaluate the
expression in \reff{start2} that includes $d-q$ sums and $q$
integrals. Our procedure will be to express \reff{start2} as an
expansion over a function $\Lambda$ and then use the Euler-Maclaurin
formula to evaluate this function. Define the following short-hand
form for a $j\!-\!q$ dimensional sum over $q$ integrals: \beq \sumjq
\!\!\int^q \equiv  \sum_{\substack{n_i=-\infty\\i=1,\ldots,
j-q}}^{\infty} \int_{-\infty}^{\infty}\expo{-\lambda \,\sqrt{n_1^2 +
\cdots + n_{j-q}^2 +x_1^2+\cdots+x_q^2}}\,\,dx_1 \ldots dx_q\,.
\eeq{multiple2} where $j$ runs from $q$ to $d-1$ (the case $j=q$
corresponds to no sums, only $q$ integrals). The reader may wonder
why we chose a definition with $j-q$ sums instead of just simply
$j$. The reason is that the total number of sums plus integrals is
then $j$ and this simplifies things later on. We define a function
$\Lambda$ by adding one more sum to the above definition: \beq
\begin{split}
\Lambda_j(q,\lambda) &\equiv \sumprime\sumjq\!\int^q\\
&=\sumprime_{n=-\infty}^{\infty}\,\sum_{\substack{n_i=-\infty\\i=1,\ldots,j-q}}^{\infty}
\int_{-\infty}^{\infty}\expo{-\lambda\,\sqrt{n^2+n_1^2 + \cdots + n_{j-q}^2 +x_1^2+\cdots+x_q^2}}\,\,dx_1 \ldots dx_q\,
\end{split}
\eeq{j12}
where the last sum over $n$ excludes
zero. With these definitions, we make the following useful expansion of \reff{start2}:
\beq
\begin{split}
&\sum_{\substack{n_i=-\infty\\i=1,\ldots,d-q}}^{\infty} \int_{-\infty}^{\infty}\expo{-\lambda\,\sqrt{n_1^2 + \cdots +
n_{d-q}^2 + x_1^2 + \cdots + x_q^2}}\,dx_1 \ldots dx_q\\
&= \int^q + \sumprime\int^q + \sumprime\,\sum\int^q + \sumprime\,\sumtwo\int^q +\cdots +
\sumprime\,\sumdq\int^q \\
&= \int^q+ \sum_{j=q}^{d-1} \sumprime\,\sumjq\int^q = \int^q+ \sum_{j=q}^{d-1} \Lambda_j(q,\lambda)\,.
\end{split}
\eeq{testr2}
Substituting \reff{testr2} into \reff{start2} yields the regularized Casimir energy:
\beq
E^{reg}_p(q,\lambda) =
-\dfrac{\pi\,v}{L^{q+1}}\,\prod_{i=1}^q L_i\,\,\left(\partial_{\lambda} \int^q +\sum_{j=q}^{d-1}
\partial_{\lambda}\Lambda_j(q,\lambda)\right)\,.
\eeq{def2} In the above expression, we need to separate the
divergent part due to the continuum from the finite part related to
the Casimir energy as $\lambda\!\to\!0$. The term
$\partial_{\lambda} \int^q$ contains no sums, only multiple
integrals. It is immediately clear that this term contributes purely
a continuum divergent part as $\lambda\to 0$ and hence makes no
contribution to the finite Casimir energy. We now need to find an
expression for $\Lambda_j(q,\lambda)$ given by \reff{j12} and
extract the finite part related to it. To this end we apply the
Euler-Maclaurin formula that converts sums to integrals. The
Euler-Maclaurin formula is given by \cite{Arfken}: \beq
\sum_{n=1}^{\infty} f(n) =\int_0^\infty f(x)\,dx -
\dfrac{1}{2}\,f(0) -
\sum_{p=1}^{s}\dfrac{1}{(2p)!}\,B_{2p}\,f^{(2p-1)}(0) + R_s
\eeq{euler} where $f^{(2p-1)}(0)$ are odd derivatives evaluated at
zero and $s$ is a positive integer. The form above for the
Euler-Maclaurin formula assumes that the function $f(n)$ and its
derivatives are zero at infinity. $R_s$ is the remainder term given
by \cite{Arfken} \beq R_s= -\dfrac{1}{(2s)!} \int_0^1\, B_{2s}(x)
\,\sum_{\nu=0}^{\infty}\, f^{2s}\,(x+\nu)\,dx \eeq{remainder} where
$B_{2s}(x)$ are Bernoulli functions and $ f^{2s}\,(x+\nu)$ are even
derivatives of $f$ with respect to $x$.

In applying the Euler-Maclaurin formula to determine
$\Lambda_j(q,\lambda)$, the function $f$ in question is the
exponential function appearing in \reff{j12}. Regardless of the
value of $p$, this exponential function has the property that
$f^{2p-1}(0)$ is zero for all sums in \reff{j12} except the last one
over $n$. A proof of this is given in the appendix of \cite{Ariel}.
If $f^{2p-1}(0)$ is zero for all $p$ it follows that the sum from
$p=1$ to $s$ in \reff{euler} is zero independent of $s$. This
implies that $R_s$ given by \reff{remainder} has the same  value for
any given $s$ for the case of our exponential function. This is
proven explicitly in the appendix of \cite{Ariel2}. For calculations
we can simply choose $s$ equal to $1$. Since $f^{2p-1}(0)$ is zero
for all sums except the last one, the Euler-Maclaurin formula for
those sums reduces to \beq \sum_{n=1}^{\infty} f(n) =\int_0^\infty
f(x)\,dx - \dfrac{1}{2}\,f(0) -\dfrac{1}{2} \int_0^1\, B_2(x)
\,\sum_{\nu=0}^{\infty}\, \dfrac{d^2}{dx^2} f(x+\nu)\,dx
\eeq{Euler2} where $B_2(x)=x^2-x+1/6$. The function $f$ in
\reff{j12} has the property $f(n_i)=f(-n_i)$. The sum over a given
$n_i$ can therefore be written as \beq
\begin{split}
&\sum_{n_i=-\infty}^{\infty} f(n_i) = 2\,\sum_{n_i=1}^{\infty}\,f(n_i) + f(0)\\
&=2 \Biggl(\int_0^\infty \!\!\!\!f(x)\,dx - \dfrac{1}{2}\,f(0)-\dfrac{1}{2} \int_0^1\! B_2(x) \,\sum_{\nu=0}^{\infty}\, \dfrac{d^2}{dx^2} f(x+\nu)\,dx \Biggr) + f(0)\\
&=\int_{-\infty}^{\infty} f(x)\,dx - R\\
\end{split}
\eeq{firstsum}
where $R$ is a remainder given by
\beq
R = \int_0^1\,B_2(x) \,\sum_{\nu=0}^{\infty}\,
\dfrac{d^2}{dx^2} f(x+\nu)\,dx \,.
\eeq{R}
From \reff{firstsum} we see that each sum in \reff{j12}, except the
last one, can be replaced by an integral minus $R$. We therefore have the operator prescription $\sum \to \int -
R$. Applying the operator $j-q$ times and then inserting the result in \reff{j12} yields
\beq
\sumjq = \Biggl(\int -R\Biggr)^{j-q} = \int^{j-q} + \sum_{m=1}^{j-q} (-1)^m
\binom{j-q}{m} \int^{j-q-m} R^m
\eeq{intr}
and
\beq
\begin{split}
\Lambda_j(q,\lambda) &\equiv \sumprime\sumjq\int^q = \sumprime\int^{j} + \sumprime\sum_{m=1}^{j-q} (-1)^m \binom{j-q}{m} \int^{j-m}
R^m
\\&=2^{j+1}\,\sum_{n=1}^{\infty}\,\int_{0}^{\infty}\expo{-\lambda\,\sqrt{n^2+ x_1^2 + \cdots + x_{j}^2}}\, dx_1\ldots dx_{j} + R_j(q,\lambda)
\end{split}
\eeq{Lambda7}
where $R_j(q,\lambda)$ is a remainder given by
\beq
R_j(q,\lambda)\equiv \sum_{m=1}^{j-q}
\sum_{n=1}^{\infty} (-1)^m \,2\,\binom{j\!-\!q}{m} \int^{j-m} R^m \,.
\eeq{Rjlam2}
Substituting $R$ given by \reff{R}
into \reff{Rjlam2} yields
\beq
\begin{split}
&R_j(q,\lambda)=\sum_{m=1}^{j-q} \sum_{n=1}^{\infty}(-1)^m \bino{j-q}{m}\,2^{j-m+1}
\int_{0}^{\infty}\,\int_0^1\,\prod_{i=1}^{m}\sum_{\nu_i=0}^{\infty} B_2(x_i) \,\dfrac{\partial^2}{\partial
x_i}\\& \quad\expo{-\lambda\,\sqrt{n^2+ (x_1+\nu_1)^2 + \cdots + (x_m+\nu_m)^2 + y_1^2 + \cdots + y_{j-m}^2}}\,
dx_1\ldots dx_m\,dy_1\ldots dy_{j-m}
\end{split}
\eeq{Rjlambda5}
where the integrations from  $0$ to $1$ and $0$ to $\infty$ are over the $x$'s and
$y$'s respectively. The function $\Lambda$ given by \reff{Lambda7} contains two terms. The first term leads to the analytical part and the
second term $R_j(q,\lambda)$ yields the remainder. In the limit $\lambda=0$, $R_j(q,\lambda)$ is zero but not its
derivative with respect to $\lambda$. It is the derivative with respect to $\lambda$ that enters into the
Casimir energy \reff{def2}. There is therefore a non-zero contribution to the Casimir energy coming from the
remainder term and we fully evaluate it later on. For now, let us evaluate the analytical term in \reff{Lambda7}. It can be reduced to an infinite sum over
the modified Bessel function $K_0(\lambda\,n)$ which has a useful series expansion. We first note that the
integral in \reff{Lambda7} can be expressed in terms of the modified Bessel function $K_{\frac{j-1}{2}}(\lambda\,n)$
\cite{Gradshteyn}:
\beq
\begin{split}
&\int_0^{\infty} \expo{-\lambda\,\sqrt{n^2+ x_1^2 + \cdots + x_{j}^2}}\, dx_1\ldots dx_{j}\\
& = -2^{\frac{1-j}{2}}\,\pi^{\frac{j-1}{2}}\,\dfrac{d}{d\lambda}\Bigl(K_{\frac{j-1}{2}}(\lambda \,n)\,
\Bigl(\frac{n}{\lambda}\Bigr)^{\frac{j-1}{2}}\Bigr)\,.
\end{split}
\eeq{in}
The modified Bessel function $K_{\frac{j-1}{2}}(\lambda\,n)$ can be expressed as multiple derivatives
of $K_0(\lambda\,n)$ \cite{Gradshteyn}:
\beq
K_{\frac{j-1}{2}}(\lambda\,n)\,\Bigl(\frac{n}{\lambda}\Bigr)^{\frac{j-1}{2}} = (-1)^{\frac{1-j}{2}}
\Bigl(\frac{d}{\lambda\,d\lambda}\Bigr)^{\frac{j-1}{2}} K_0(\lambda\,n)\,.
\eeq{bess}
Substituting \reff{bess}
and \reff{in} into \reff{Lambda7} yields $\Lambda_j(q,\lambda)$ as an infinite sum over the modified Bessel function
$K_0(\lambda\,n)\,$:
\beq
\Lambda_j(q,\lambda) =  2^{\frac{j+3}{2}}\,\pi^{\frac{j-1}{2}}
(-1)^{\frac{3-j}{2}}\dfrac{d}{d\lambda} \Bigl(\frac{d}{\lambda\,d\lambda}\Bigr)^{\frac{j-1}{2}}
\sum_{n=1}^\infty K_0(\lambda\,n) + R_j(q,\lambda).
\eeq{sumko}
The infinite sum over the modified Bessel function
$K_0(\lambda\,n)$ has the following series expansion \cite{Gradshteyn}:
\beq
\sum_{n=1}^\infty K_0(\lambda\,n) =
\dfrac{1}{2}\left\{C + \ln(\lambda/4\pi)\right\} + \dfrac{\pi}{2\,\lambda} + \pi \sum_{m=1}^\infty
\left(\dfrac{1}{\sqrt{\lambda^2 + 4\,m^2\,\pi^2}} - \dfrac{1}{2\,m\,\pi}\right)\,.
\eeq{series}
By substituting
\reff{series} into \reff{sumko} we obtain $\Lambda_j(q,\lambda)$ as an analytic expression plus the remainder
$R_j(q,\lambda)$:
\beq
\begin{split}
\Lambda_j(q,\lambda) &= -\dfrac{1}{\lambda^j}\,2^j \,\pi^{\frac{j-1}{2}}\, \Gamma(\tfrac{j+1}{2}) +
\dfrac{1}{\lambda^{j+1}}\,2^{j+1}\,\pi^{\frac{j}{2}}\,\Gamma(\tfrac{j+2}{2}) \\
&\qquad\qquad\qquad\qquad+ \lambda
\,\,2^{j+2}\,\,\Gamma(\tfrac{j+2}{2})\,\pi^{\frac{j}{2}}\,\chi_j(\lambda) + R_j(q,\lambda)\\
\end{split}
\eeq{finito}
\beq \word{where} \chi_j(\lambda) \equiv \sum_{m=1}^\infty \dfrac{1}{(\lambda^2 +
4\,m^2\,\pi^2)^{\frac{j+2}{2}}} \,\,.
\eeq{chi}
To obtain the regularized vacuum energy $E^{reg}_p(q,\lambda)$
given by \reff{def2} we need to evaluate the derivative of $\Lambda$:
\beq
\begin{split}
\partial_{\lambda} \Lambda_j(q,\lambda) &= \dfrac{j}{\lambda^{j+1}}\,2^j \,\pi^{\frac{j-1}{2}}\, \Gamma(\tfrac{j+1}{2}) -\dfrac{j+1}{\lambda^{j+2}}\,2^{j+1}\,\pi^{\frac{j}{2}}\,\Gamma(\tfrac{j+2}{2}) \\
& \qquad+ \,2^{j+2}\,\,\Gamma(\tfrac{j+2}{2})\,\pi^{\frac{j}{2}}\,\chi_j(\lambda)
+\lambda\,\,2^{j+2}\,\,\Gamma(\tfrac{j+2}{2})\,\pi^{\frac{j}{2}}\,\partial_{\lambda} \,\chi_j(\lambda)\\&
 \qquad\qquad+\,\partial_{\lambda} R_j(q,\lambda)\,.
\end{split}
\eeq{derlambda}
We now take the limit as $\lambda \to 0$ in \reff{derlambda}. Note that the first two terms in
\reff{derlambda} are divergent in this limit and represent the infinite continuum energy of surface and volume
terms respectively. The Casimir energy is the difference between the discrete and continuum case and therefore these two
terms need to be subtracted out. We therefore define
\beq
\partial_{\lambda} \Lambda^{finite}_j(q,\lambda) = \,2^{j+2}\,\,\Gamma(\tfrac{j+2}{2})\,\pi^{\frac{j}{2}}\,\chi_j(\lambda) +\lambda\,\,2^{j+2}\,\,\Gamma(\tfrac{j+2}{2})\,\pi^{\frac{j}{2}}\,\partial_{\lambda} \,\chi_j(\lambda) +
\partial_{\lambda} R_j(q,\lambda).
\eeq{finitelam}
 The above terms in the limit $\lambda=0$ are
 \beq
\lim_{\lambda \to 0} \chi_j(\lambda) =  (2\,\pi)^{-j-2} \zeta(j+2)\quad; \quad\lim_{\lambda \to 0}\,
\partial_{\lambda} \,\chi_j(\lambda) =0
\eeq{limitlambda}
and we define $R_j(q)$ as
\beq
 R_j(q) \equiv \lim_{\lambda\to 0}\, \partial_{\lambda} \,R_j(q,\lambda)\,.
\eeq{val}
Substituting \reff{limitlambda} and \reff{val} into \reff{finitelam} we obtain the compact form
\beq
\begin{split}
\lim_{\lambda \to 0} \partial_{\lambda} \Lambda^{finite}_j(q,\lambda) =
\Gamma(\tfrac{j+2}{2})\,\pi^{\frac{-j-4}{2}}\, \zeta(j+2) + R_j(q) \,.
\end{split}
\eeq{lambdafine} $R_j(q)$ is the multidimensional remainder which
contributes to the Casimir energy. This is evaluated in Appendix A
and the result is: \beq R_j(q) = \dfrac{1}{\pi} \sum_{m=1}^{j-q}
2^{\,m+1}\,\binom{j-q}{m}\,
\sum_{n=1}^{\infty}\sum_{\ell_{1,\ldots,m}=1}^{\infty}
\dfrac{n^{\frac{j+1}{2}}\,K_{\frac{j+1}{2}}(2\,\pi\,n\,\sqrt{\ell_1^2
+\cdots+\ell_m^2})}{(\ell_1^2 +\cdots+
\ell_m^2)^{\tfrac{j+1}{4}}}\,. \eeq{rjfinal} Note that $R_j(q)$ is
zero for $j=q$.  The above expression \reff{rjfinal} for the
remainder is highly convenient. First, it converges rapidly. The
Bessel functions decrease rapidly and therefore only the very first
few numbers in each sum are needed to reach high accuracy. Secondly,
clever algorithms for Bessel functions are well incorporated in many
software packages making numerical computation of the remainder easy
and accurate. The finite part of \reff{def2} in the limit
$\lambda=0$ yields the Casimir energy for the periodic case: \beq
\begin{split}
E_p(q,d) &= -\dfrac{\pi\,v}{L^{q+1}}\,\prod_{i=1}^q L_i \,\,\sum_{j=q}^{d-1} \lim_{\lambda \to 0}
\partial_{\lambda} \Lambda^{finite}_j(q,\lambda)\\
&= -\dfrac{\pi\,v}{L^{q+1}}\prod_{i=1}^q L_i \,\sum_{j=q}^{d-1} \Gamma(\tfrac{j+2}{2})\,\pi^{\frac{-j-4}{2}}\,
\zeta(j+2)+ R_{j}(q)
\end{split}
\eeq{periodicfinal2} with $R_j(q)$ given by \reff{rjfinal}. Equation
\reff{periodicfinal2} is the Casimir energy of a massless scalar
field moving with velocity $v$ in a $d$-dimensional rectangular box
with periodic boundary conditions where $d-q$ sides have length $L$
and $q$ sides have much larger lengths. Note the convenient break-up
into two terms: a finite analytical formula over the well-known
Riemann zeta and gamma functions plus a remainder. Since $R_j(q)$ is
zero for $j=q$, the sum for the remainder starts at $j=q+1$ and is
therefore non-zero only if $d\ge q + 2$ i.e. non-zero only if there
is at least two small dimensions on top of the $q$ large dimensions.

We can now readily express the Casimir energies for the Neumann and
Dirichlet cases as sums over the periodic ones. In \reff{vacuum}, the sums for the periodic case start at $-\infty$,
while for Neumann and Dirichlet cases they start at $0$ and $1$ respectively. We can express the sums from $0$ or
$1$ to $\infty$ in terms of sums from $-\infty$ to $\infty$. The functions being summed have the property $f(n)
= f(-n)$. We therefore have the relation $\sum_{0}^{\infty} f(n) =
\tfrac{1}{2}\,\sum_{-\infty}^{\infty} f(n) + \tfrac{1}{2} \,f(0)$ which can be expressed as an operator
$\sum_{0}^{\infty} \to \tfrac{1}{2} \bigl( \sum_{-\infty}^{\infty} + 1\bigr)$. Applying the operator $d-q$ times
yields:
\beq
\begin{split}
E_N(q,d)&\equiv \dfrac{\pi\,v}{2\,L^{q+1}}\,\prod_{i=1}^q L_i \int_0^{\infty}\,\Bigl(\sum_{0}^{\infty}\Bigr)^{d-q} \to \dfrac{\pi\,v}{2\,L^{q+1}}\,\prod_{i=1}^q L_i \dfrac{1}{2^d} \int_{-\infty}^{\infty}\Bigl( 1+ \sum_{-\infty}^{\infty} \Bigr)^{d-q} \\&=
2^{-d-1}\,\dfrac{\pi\,v}{L^{q+1}}\,\prod_{i=1}^q L_i  \sum_{m=1}^{d-q} \bino{d-q}{m} \int^q \Bigl(\sum_{-\infty}^{\infty}\Bigr)^m \\&= 2^{-d-1}\, \sum_{m=1}^{d-q}
\bino{d-q}{m} \,E_p(q,q+m)
\end{split}
\eeq{operator2}
Substituting \reff{periodicfinal2} into \reff{operator2} yields the Neumann Casimir energy:
\beq
\begin{split}
E_N(q,d)&= -2^{-d-1}\,\dfrac{\pi\,v}{L^{q+1}}\prod_{i=1}^q L_i\\& \,\sum_{j=q}^{d-1} \sum_{m=j-q+1}^{d-q} \bino{d-q}{m} \left(\,\Gamma(\tfrac{j+2}{2})\,\pi^{\frac{-j-4}{2}}\,
\zeta(j+2) + R_j(q)\,\right).
\end{split}
\eeq{Neumann2}
For the Dirichlet case, $\sum_{1}^{\infty} f(n) = \tfrac{1}{2}\,\sum_{-\infty}^{\infty} f(n) -
\tfrac{1}{2} \,f(0)$ and we obtain
\beq
E_D(q,d)=2^{-d-1} \sum_{m=1}^{d-q} (-1)^{d-q+m} \bino{d-q}{m} \, E_p(q,q+m)\,.
\eeq{Dirichlet3}
Substituting \reff{periodicfinal2} into \reff{Dirichlet3} yields the Dirichlet Casimir energy:
\beq
\begin{split}
E_D(q,d)&=2^{-d-1}\,\dfrac{\pi\,v}{L^{q+1}}\prod_{i=1}^q L_i \\& \sum_{j=q}^{d-1} (-1)^{d+j} \bino{d\!-\!q\!-\!1}{j\!-\!q}
\left(\,\Gamma(\tfrac{j+2}{2})\,\pi^{\frac{-j-4}{2}}\,\zeta(j+2)+ R_j(q)\,\right).
\end{split}
\eeq{Dirich2}

A special case is that of Dirichlet conditions for parallel plates
where all sides except one are large i.e. $q=d-1$. $R_j(q)$ is then
zero and only $j=d-1$ is summed: \beq
E_{||}(d)=-2^{-d-1}\,\dfrac{\pi\,v}{L^d}\prod_{i=1}^{d-1} L_i \,\,
\Gamma(\tfrac{d+1}{2})\,\pi^{\frac{-d-3}{2}}\,\zeta(d+1)
\eeq{parallel3} The Casimir pressure for the parallel plates is
then: \beq P_{||}(d) \equiv -\dfrac{\partial E_{||}}{\partial V}
=-\dfrac{\hbar\,v\,d}{(2\,L)^{\,d+1}}\,
\Gamma(\tfrac{d+1}{2})\,\pi^{\frac{-d-1}{2}}\,\zeta(d+1) \eeq{parap}
where $V$ is the volume $L \prod_{i=1}^{d-1} L_i$ of the parallel
plates and we have re-inserted $\hbar$. The result \reff{parap} is
in agreement with the higher-dimensional parallel plate cut-off
calculation of \cite{Svaiter2} if we set $v$ and $L$ to unity. For
three dimensions we set $d=3$ and obtain: \beq P_{||}(3)
=-\dfrac{\pi^2}{480}\,\dfrac{\hbar\,v}{L^4} \eeq{parap2} where we
used the fact that $\zeta(4)=\pi^4/90$. This result is in agreement
with the Casimir calculation for quantum fluctuations in a dilute
Bose-Einstein condensate at zero temperature that was recently
carried out by \cite{Roberts2,Roberts}. As previously mentioned,
though the BEC has a non-linear dispersion relation the Casimir
energy only picks out the low frequency part since the higher
frequencies act as a continuum. The low frequency part is linear and
the dispersion relation is equivalent to that of a massless
Klein-Gordon field with speed of light replaced by speed of sound.
The pressure in \reff{parap2} is negative implying attraction and
decreases to the fourth power of the distance as in the
electromagnetic case. In fact, the classic electromagnetic result
$-\tfrac{\pi^2}{240}\,\tfrac{\hbar}{L^4}$ for parallel-plates can be
obtained by multiplying \reff{parap2} by $2$ for two polarizations
and setting $v$ equal to $1$ for the speed of light.

Equations \reff{periodicfinal2}, \reff{Neumann2} and \reff{Dirich2}
for the Casimir energies contain products of the large dimensions
$L_i$ which can be arbitrarily large. It is of more physical
interest to obtain the energy densities $\varepsilon$ which depend
on $L$ only. Dividing the Casimir energies by the volume
$V=L^{d-q}\,\prod_{i=1}^q L_i$ yields \beq
\begin{split} \varepsilon_p&= -\dfrac{\pi\,v}{L^{d+1}} \sum_{j=q}^{d-1} \Gamma(\tfrac{j+2}{2})\,\pi^{\frac{-j-4}{2}}\,
\zeta(j\!+\!2)+ R_{j}(q)\\\varepsilon_N&= -\dfrac{\pi\,v}{(2\,L)^{\,d+1}}\sum_{j=q}^{d-1} \sum_{m=j-q+1}^{d-q} \binom{d\!-\!q}{m} \left(\,\Gamma(\tfrac{j+2}{2})\,\pi^{\frac{-j-4}{2}}\,
\zeta(j\!+\!2) + R_j(q)\,\right)\\\varepsilon_D&= \dfrac{\pi\,v}{(2\,L)^{\,d+1}}\sum_{j=q}^{d-1} (-1)^{d+j} \,\binom{d \!-\! q \!-\! 1}{j-q}
\left(\,\Gamma(\tfrac{j+2}{2})\,\pi^{\frac{-j-4}{2}}\,\zeta(j\!+\!2)+ R_j(q)\,\right).
\end{split}
\eeq{periodicfinal3} The three equations in \reff{periodicfinal3}
are our final results for the periodic, Neumann and Dirichlet
Casimir energy densities for massless scalar fields moving with wave
velocity $v$ in a $d$-dimensional rectangular cavity where  $d-q$
sides have equal length $L$ and $q$ sides have much larger length.
The expressions contain a dominant finite analytical part plus a
fast-converging remainder $R_j(q)$ given by \reff{rjfinal}. General
formulas for arbitrary lengths are obtained in appendix B.

\section{Physical interpretation of Casimir energy
formulas}

The Casimir energy formula  \reff{periodicfinal2} for periodic
boundary conditions and \reff{Neumann2} and \reff{Dirich2} for
Neumann and Dirichlet conditions respectively have a clear physical
picture or interpretation. Excluding the remainder, the formulas can
be viewed as the energy needed to set up the parallel plates from
which the rectangular cavity is constructed. For example, consider
the case $d=3$ and $q=0$ corresponding to a cube (hypertorus for
periodic) with sides of length $L$. The cube is built out of three
sets of parallel plates. In \reff{periodicfinal2} this corresponds
to summing the term $\Gamma(\tfrac{j+2}{2})\,\pi^{\frac{-j-4}{2}}\,
\zeta(j\!+\!2)$ for $j=0,1$ and $2$. To build the cube, one begins
by placing two plates a distance $L$ apart. This corresponds to
$j=2$. Adding two more plates corresponds to $j=1$ and the last two
plates completes the cube and corresponds to $j=0$. We now show
mathematically that the Casimir energy is the sum of parallel plate
energies plus a remainder. Consider periodic boundary conditions.
The energy for parallel plates defined by letting $q=d-1$ in
\reff{periodicfinal2} is: \beq E_{p_{\,\,||}}(d) =
-\dfrac{\pi\,v}{L^d}\prod_{i=1}^{d-1} L_i
\,\,\,\Gamma(\tfrac{d+1}{2})\,\pi^{\frac{-d-3}{2}}\,\zeta(d+1)\,.
\eeq{prs} $R_j(q)$ is zero for parallel plates and this is why it is
not present in \reff{prs}. The parallel plate energy in $j+1$
dimensions is \beq E_{p_{\,\,||}}(j+1) =
-\dfrac{\pi\,v}{L^{j+1}}\prod_{i=1}^{j} L_i
\,\,\,\Gamma(\tfrac{j+2}{2})\,\pi^{\frac{-j-4}{2}}\,\zeta(j+2)\,.
\eeq{prs2} In \reff{periodicfinal2}, $j \ge q$. Therefore the first
$q$ products in $\prod_{i=1}^{j} L_i$ are large and the rest are
equal to $L$ so that the above product $\prod_{i=1}^{j} L_i$ can be
replaced by $L^{j-q}\prod_{i=1}^{q} L_i$ yielding \beq
E_{p_{\,\,||}}(j+1) = -\dfrac{\pi\,v}{L^{q+1}}\prod_{i=1}^q L_i
\,\,\,\Gamma(\tfrac{j+2}{2})\,\pi^{\frac{-j-4}{2}}\,\zeta(j+2)\,.
\eeq{periodicparallel} Substituting \reff{periodicparallel} in
\reff{periodicfinal2} yields: \beq E_p(q,d)=\sum_{j=q}^{d-1}
\left(\,E_{p_{\,\,||}}(j+1) + R_j(q)\,\right) \eeq{pp2} As can be
seen, the Casimir energy in a $d$-dimensional space with $q$ large
dimensions is the sum of parallel plates immersed in different
dimensions plus a remainder. When building the rectangular cavity
out of successive parallel plates, the first parallel plates have
$d-1$ large dimensions, the second have $d-2$ large dimensions and
so on until the last set which has $q$ large dimensions. In short,
the $d-q$ dimensional resonator is the sum of one-dimensional
resonators each immersed in a different dimension ranging from $q+1$
to $d-1$.

What is the physical interpretation for the remainder? The energy
for parallel plates are by definition those for isolated plates in
vacuum. However, to construct the rectangular cavity, one adds
plates to other plates already present. To clarify this difference
consider two scenarios. Scenario I: plates are brought together in
vacuum in a two dimensional space. This leaves one dimension which
is large. Scenario II: consider a three dimensional space where
there is already a pair of parallel plates. Now add another pair of
plates. This leaves one dimension which is large as in scenario I.
The main point is this: the energy in scenario II for adding the
second set of plates is almost but not exactly equal to the energy
of the plates in scenario I. The reason is that in scenario II there
is also an interaction energy due to the presence of the other
plates. The remainder term is therefore an `interaction' or
potential energy arising from the nonlinearity of the energy when
waves moving along different directions are added. By interaction
energy we do not mean that there is a Feynman diagram where scalar
fields meet at a vertex. That would be a nonlinear theory like
$\lambda \,\phi^4$. What we have here is a linear theory and the
waves obey the superposition principle. However, the energy is
clearly not linear. This is reminiscent of what occurs in classical
electrodynamics. In vacuum, the theory is linear and one can add two
electric field vectors but the energy itself is not linear since it
is proportional to the square of the electric field. What we usually
call the potential energy between two static charges $q_1$ and $q_2$
is nothing but the interaction energy between the electric field
${\bf E_1}$ produced by the first charge and the electric field
${\bf E_2}$ produced by the second charge. The energy density is
proportional to $({\bf E_1 + E_2})^{\,2} = E_1^{\,2} + E_2^{\,2} + 2
\,{\bf E_1\cdot E_2}$ and the integration of the cross-term $2
\,{\bf E_1\cdot E_2}$ over all space yields the well-known potential
energy proportional to $q_1\,q_2/r$ where $r$ is the distance
between the charges. The remainder term is similarly a potential
energy arising from the nonlinearity of the energy.

We can now make predictions about the behaviour of the remainder for
periodic, Neumann and Dirichlet boundary conditions. We predict the
following:
\begin{itemize}
\item percentage wise, the periodic case will have the highest remainder,
the Dirichlet case the smallest, and Neumann in between
\item the remainder grows with the space dimension for the
periodic and Neumann cases but actually decreases for the Dirichlet
case
\end{itemize}

Let us see how we can make such predictions. The Casimir energy is
the difference between discrete and continuum modes. As the
frequency increases the discrete approaches the continuum. Therefore
the Casimir energy picks out the low frequency or low energy
behaviour. Moreover, the lower the energy, the more nonlinear is the
change in energy. Higher energies are closer to the continuum and
changes are more linear. The minimum energy mode for the periodic
and Neumann cases is zero (the case when all $n_i$'s are zero). For
Dirichlet the minimum energy mode occurs when all $n_i$'s are equal
to $1$. For concreteness let the space dimension be $5$. For
periodic and Neumann the smallest nonzero energy state occurs when
one $n_i$ is $1$ so that one of five slots is filled with $1$ e.g.
(0,1,0,0,0) while for Dirichlet the minimum energy starts at
(1,1,1,1,1). Now add $1$ to both cases (creating states with two
$1$'s like (0,1,0,0,1) and states like (1,2,1,1,1)). The percentage
change in the energy in the Dirichlet case will not be large because
the energy started off large. The energy changes almost linearly
leading to a small remainder. As the dimension increases, the energy
for the Dirichlet case starts off even higher and the change is even
less. For Dirichlet, we therefore predict the remainder to be a very
small percentage of the energy and that it {\it decreases} as the
space dimension grows. In the periodic and Neumann case, the energy
starts off low, so the change is a larger percentage of the initial
energy and therefore more nonlinear than in the Dirichlet case. This
effect is greatly accentuated by the fact that are many more
low-energy combinations for the Neumann and periodic case compared
to the Dirichlet case. For example, there are $5$ ways to place $2$
in (1,2,1,1,1) but there are $10$ ways to arrange the two $1$'s in
(0,1,0,0,1). The remainder will therefore be considerably larger in
the Neumann and periodic case. Moreover, the remainder for periodic
and Neumann cases will grow as the dimension increases because as
the number of zeros increases there are simply more possible
low-energy combinations and this increases the nonlinear effect.
Finally, the periodic case has the largest remainder of all the
cases because negative $n$'s are allowed, so that in our state
(0,1,0,0,1) one can also have combinations with $-1$ leading to
considerably more low-energy contributions than in the Neumann case.
Our numerical results confirm all these trends.

\section{ Epstein zeta in odd dimensions as products of
one-dimensional sums plus remainder}

When applied to a rectangular geometry, the zeta function
regularization technique via analytical continuation yields quickly
a finite expression for the Casimir energy in terms of homogeneous
Epstein zeta functions. The subtraction of two infinities does not
explicitly appear anywhere in the process. This is a great advantage
over the cut-off technique. We use zeta function regularization here
to obtain quickly an expression for the Casimir energy in terms of
Epstein zeta functions for the periodic case. Our main goal however
is to express the homogeneous Epstein zeta function for $3$,$5$ and
$7$ dimensions in terms of products of one-dimensional sums plus a
small remainder. Readers interested in getting a deeper
understanding of the zeta regularization technique as well as other
techniques such as heat-kernel methods are referred to the following
books \cite{book1,book2,book3}. A sample of older and more recent
articles where these techniques are applied in various contexts
ranging from gravitation to condensed-matter can be found in
\cite{Kirsten1,Elizalde3,Elizalde4,Fulling,
Elizalde5,Kirsten2,Schakel,Xin,Ortenzi,
E1,E2,Elizalde6,Cognola,Kirsten3}. For the case of rectangular
cavities in arbitrary $d$ dimensions these techniques have been
applied in \cite{Wolfram,Neto,Li}.

Though one can compute a finite numerical result, extra work must be
done to express the Epstein zeta function in a compact analytical
form. Define the Epstein zeta function $Z_d(a_1,\ldots,a_d;s)$ as:
\beq Z_d(a_1,\ldots,a_d;s)\equiv
\sideset{}{'}\sum_{\substack{n_i=-\infty\\i=1,\ldots,
d}}^{\infty}\left[(a_1\,n_1)^2 + \cdots + (a_d\,n_d)^2\right]^{-s}
\eeq{Epstein} where the prime excludes the case where all $n$'s are
zero and absolute convergence requires Re $ s > d/2$. {\it Our
definition differs from the standard one by a factor of 2 in the
power i.e. we have $-s$ instead of $-s/2$}. This definition is
chosen as it simplifies our final expressions. We focus on the case
of the hypercube, where all the $a$'s are equal and can be pulled
out of the sum in \reff{Epstein} (for simplicity we set them to
unity). This yields the homogeneous Epstein zeta function $Z_d(s)$.
The vacuum energy in $d$ dimensions for periodic boundary conditions
is trivial to write in terms of $Z_d(s)$: \beq
\begin{split} E^{vac}_{p}(0,d)&= \dfrac{\pi
\,v}{L}\sum_{\substack{n_i=-\infty\\i=1,\ldots, d}}^{\infty} (n_1^2
+ \cdots + n_d^2)^{1/2}\\&=\dfrac{\pi \,v}{L}
\,Z_d(-1/2)\,.\end{split} \eeq{periodiczeta} Now $Z_d(-1/2)$ is
formally infinite if \reff{Epstein} is applied in a straightforward
fashion. It therefore requires regularization. The keystone of the
zeta regularization technique is analytic continuation and the
existence of a reflection formula. Like the Riemann zeta function,
the Epstein zeta function has an integral representation which
yields an analytic continuation over the entire complex plane except
for a pole at $s=d/2$. The representation leads to the following
functional relation or reflection formula: \beq
\pi^{-s}\,\Gamma(s)\,Z_d(s)=\pi^{s-d/2}\,\Gamma(d/2-s)\,Z_d(d/2-s)\,.
\eeq{reflection} We therefore obtain that \beq Z_d(-1/2) =
-0.5\,Z_d(\tfrac{d+1}{2})\,\Gamma(\tfrac{d+1}{2})\,\pi^{\tfrac{-3-d}{2}}\eeq{recursion}
and \reff{periodiczeta} reduces to the Casimir energy
 \beq
E_{p}(0,d)=\dfrac{-\pi\,v}{2\,L}\,Z_d(\tfrac{d+1}{2})\,\Gamma(\tfrac{d+1}{2})\,\pi^{\tfrac{-3-d}{2}}
\eeq{periodzet} Clearly, the Casimir energy is finite since
$Z_d(\tfrac{d+1}{2})$ converges. The reader should appreciate just
how quickly the zeta function technique yields this result.

The homogeneous Epstein zeta function $Z_d(s)$ can be expressed in
terms of sums over the arithmetical function $r_d(n)$ which is the
number of representations of an integer $n$ as a sum of $d$ squares
without regard to sign or order: \beq
\begin{split}
Z_d(s)&\equiv \sideset{}{'}\sum_{\substack{n_i=-\infty\\i=1,\ldots, d}}^{\infty}\left[n_1^2 + \cdots + n_d^2\right]^{-s}\\
&=\sum_{n=1}^{\infty} \dfrac{r_d(n)}{n^s}\,.
\end{split}
\eeq{rdn} We can therefore use results from number theory on
$r_d(n)$ to obtain directly formulas for the Epstein zeta function.
It turns out that formulas for $r_d(n)$ which are not complicated
exist in 2, 4, 6 and 8 dimensions and these can be used to obtain
the Epstein zeta function \reff{rdn} as products of one dimensional
sums with no remainder. The formula for dimension 1 is trivial (by
definition a Riemann zeta function) but formulas for 3, 5 and 7
dimensions eluded number theorists until a major breakthrough in
2002 when Goro Shimura developed a systematic way of finding
formulas for $r_d(n)$ for values of $d$ up to $8$ \cite{Goro}.
Unfortunately, the odd-dimensional formulas are much more
complicated than the even ones. However, one can develop a technique
where one obtains excellent analytical expressions plus a small
remainder for $Z_3(s),Z_5(s)$ and $Z_7(s)$. This technique makes use
of number theory results in $2$,$4$,$6$ and $8$ dimensions and the
Euler-Maclaurin formula to fill in the odd-dimensional gaps. The
remainder which is explicitly evaluated turns out small because the
odd cases are derived to a large part from the even cases. The most
important case is of course $Z_3(-\tfrac{1}{2})$ since it relates to
the realistic three-dimensional Casimir energy. We obtain a nice
compact analytical expression for $Z_3(s)$. The analytical part is
so accurate that it yields the correct Casimir energy to within a
remarkable $0.04\%$ as compared to $1.6\%$ from our cut-off
formulas.

We start by stating the number-theoretic formulas for
$r_2(n),r_4(n),r_6(n)$ and $r_8(n)$ and the known exact expressions
for $Z_1,Z_2,Z_4,Z_6$ and $Z_8$ obtained from them via \reff{rdn}.
We illustrate how to obtain $Z_d(s)$ via the number-theoretic
formulas for $r_d(n)$, something that may not be too familiar to
many Physicists. We choose $d=6$ as the example to illustrate as it
fills a gap in the table quoted in \cite{Wolfram} which contains
$Z_1,Z_2,Z_4$ and $Z_8$ but not $Z_6$. We then develop the
mathematical technique by which we obtain the odd-dimensional
homogeneous Epstein zeta functions.

\subsection{Exact expressions for even-dimensional Epstein
zeta function via $r_d(n)$}

As mentioned already, the arithmetical function $r_d(n)$ is the
number of representations of an integer $n$ as the sum of $d$
squares without regard to order or sign. The formulas for $r_d(n)$
for $d=2,4,6$ and $8$ are known and given by (a good history with
references can be found in \cite{mathworld}):
\begin{align}\label{wild}
&r_2(n)= 4 \sum_{d|n}\chi(d)& &r_4(n)=8 \sum_{\substack{d|n \\ 4 \negb d}} d \nonumber\\&r_6(n)=16\sum_{d|n}\chi(d')\,d^2-4\sum_{d|n}\chi(d)\,d^2& &r_8(n)=16\sum_{d|n}(-1)^{n+d}\,d^3
\end{align}
where $d'=n/d$ and $\chi(d)$ is the primitive Dirichlet character modulo 4 given by $\chi(d)=0$ if $d$ is even
and $\chi(d)=(-1)^{\frac{d-1}{2}}$ if $d$ is odd. We now evaluate $Z_d(s)$ for $d=6$:
\beq\begin{split}
Z_6(s)&=\sum_{n=1}^{\infty} \dfrac{r_6(n)}{n^s} \\&= 16\sum_{d'= \text{odd}}\sum_{d=1}^{\infty} \dfrac{(-1)^{\frac{d'-1}{2}} d^2}{(d'\,d)^s} \,- \,4 \sum_{d=\text{odd}}\sum_{p=1}^{\infty} \dfrac{(-1)^{\frac{d-1}{2}}\,d^2}{(d\,p)^s}\\
&=16\sum_{m=0}^{\infty}\sum_{d=1}^{\infty} \dfrac{(-1)^m}{(2m+1)^s\,d^{s-2}} \,-\, 4 \sum_{m=0}^{\infty}\sum_{p=1}^{\infty} \dfrac{(-1)^m}{(2m+1)^{s-2}\,p^s}\\&
=16\sum_{m=0}^{\infty}\dfrac{(-1)^m}{(2m+1)^s}\sum_{d=1}^{\infty}\dfrac{1}{d^{s-2}} \,-\, 4 \sum_{m=0}^{\infty}\dfrac{(-1)^m}{(2m+1)^{s-2}}\sum_{p=1}^{\infty}\dfrac{1}{p^s}\\&
=16\, \beta(s)\,\zeta(s-2) -4 \,\beta(s-2)\,\zeta(s)
\end{split}
\eeq{z6}
where $\beta(s)$ and $\zeta(s)$ are the Dirichlet beta and Riemann zeta function respectively defined by $\beta(s)\equiv  \sum_{n=0}^{\infty}(-1)^n/(2n+1)^s$ and $\zeta(s)\equiv \sum_{n=1}^{\infty}1/n^s$. We have illustrated how knowledge of the arithmetical function $r_6(n)$ leads to an exact and simple representation for the Epstein zeta function $Z_6(n)$ as a product of the one-dimensional sums $\beta(s)$ and $\zeta(s)$. The other Epstein zeta functions can be obtained in a similar fashion. We state them below together with $Z_6(s)$ \cite{Hardy}:
\beq
\begin{split}
&Z_1(s)=2\,\zeta(2\,s)\\& Z_2(s)=4\,\zeta(s)\,\beta(s)\\& Z_4(s)=8\,\zeta(s)\,\zeta(s-1)(1-4^{1-s})\\
&Z_6(s)=16\,\beta(s)\,\zeta(s\!-\!2)-4\,\beta(s\!-\!2)\,\zeta(s)\\&Z_8(s)\!=\!16\,\zeta(s)\,\zeta(s\!-\!3)(1\!-\!2^{1-s} \!+\!4^{2-s})\,.
\end{split}
\eeq{zetas}

\subsection{Analytical expressions for Epstein zeta
function $Z_d(s)$ in 3,5 and 7 dimensions}

As already mentioned, the formulas for $r_d(n)$ for $d=3,5$ and $7$
are much more complicated than the even ones and it is not easy to
use them to obtain analytical formulas for $Z_3,Z_5$ and $Z_7$. We
therefore develop a separate technique to find such expressions. The
Epstein zeta function $Z_d(s)$ defined in \reff{rdn} contains $d$
sums which begin at $-\infty$. It is convenient to define another
function $P_k(s)$ as $k$ sums which start at $1$: \beq P_k(s) \equiv
\sum_{\substack{n_i=1\\i=1,\ldots,k}}^{\infty}\left[n_1^2 + \cdots +
n_k^2\right]^{-s}\,. \eeq{Pks} We can express $P_k(s)$ as sums over
$Z_m(s)$: \beq P_k(s)= \sum_{m=1}^k
(-1)^{m+k}\,2^{-k}\bino{k}{m}\,Z_m(s)\,. \eeq{pkzm} Similarly, we
can express $Z_d(s)$ as sums over $P_k(s)$: \beq Z_d(s)=\sum_{k=1}^d
\bino{d}{k}\,2^k\,P_k(s)\,. \eeq{zkpk} It is instructive to map out
the main idea or process behind the technique we will use. Consider
the example of wanting to find expressions for $Z_3$. From
\reff{zkpk}, you would need to know $P_1$, $P_2$ and $P_3$. You can
find $P_1$ and $P_2$ from \reff{pkzm} since analytical expressions
for $Z_1$ and $Z_2$ are known. However, you do not know $P_3$. At
this point, you use the Euler-Maclaurin formula to express $P_3$ in
terms of $P_2$ plus a remainder. Again, you know $P_2$ in terms of
$Z_1$ and $Z_2$, so that you can finally express $Z_3$ in terms of
$Z_1$, $Z_2$ and a remainder and hence as an analytical part plus a
remainder. The process can be continued to find expressions for
$Z_5$ and $Z_7$ (and even $Z_9$ if one wants to but the expression
becomes cumbersome). We now develop the mathematical technique and
obtain our main equation. $Z_d(s)$ given by \reff{zkpk} can be
expanded as \beq
\begin{split}
Z_d(s)&=\sum_{k=1}^{d-1} \bino{d}{k}\,2^k\,P_k(s) + 2^d\,P_d(s)\\
&=\sum_{k=1}^{d-1} (-1)^{d+k+1}\,\bino{d}{k}\,Z_k(s) + 2^d\,P_d(s)
\end{split}
\eeq{Zn} where \reff{pkzm} was used. We now express $P_d(s)$ in
terms of $P_{d-1}(s)$ plus a remainder via the Euler-Maclaurin
formula \reff{euler}: \beq
\begin{split}
P_d(s) &= \sum_{\substack{n_i=1\\i=1,\ldots,d}}^{\infty}\left[n_1^2 + \cdots + n_d^2\right]^{-s}\\
&= \sum_{\substack{n_i=1\\i=1,\ldots,d-1}}^{\infty}
\int_0^{\infty}\dfrac{dx}{(x^2+n^2)^{\,s}} -\dfrac{1}{2\,n^{\,2s}}
\\&\qquad\qquad- \dfrac{1}{2}\sum_{\nu=0}^{\infty}\int_0^1
B_2(x)\dfrac{\partial^2}{\partial x^2}\dfrac{1}{((x+\nu)^2
+n^2)^{\,s}}\,dx
\end{split}
\eeq{pds} where \beq n^2\equiv n_1^2 + \cdots + n_{d-1}^2\,.
\eeq{asquared} The first integral in \reff{pds} can readily be
evaluated: \beq
\begin{split}
\int_0^{\infty}\dfrac{dx}{(x^2+n^2)^{\,s}} &=
\dfrac{1}{n^{2s-1}}\dfrac{\Gamma(s-\tfrac{1}{2})}{\Gamma(s)}\dfrac{\sqrt{\pi}}{2}\\&
=\dfrac{\alpha(s)}{2}\,\dfrac{1}{n^{2s-1}}
\end{split}
\eeq{int3} where $\alpha(s)$ is defined by \beq \alpha(s)\equiv
\dfrac{\sqrt{\pi}\,\,\Gamma(s-\tfrac{1}{2})}{\Gamma(s)}\,.
\eeq{asquared2} Inserting \reff{int3} into \reff{pds} yields \beq
\begin{split}
P_d(s) &= \sum_{\substack{n_i=1\\i=1,\ldots,d-1}}^{\infty}
\dfrac{\alpha(s)}{2\,n^{2s-1}} -\dfrac{1}{2\,n^{\,2s}}
\\&\qquad\qquad\qquad- \dfrac{1}{2}\sum_{\nu=0}^{\infty}\int_0^1
B_2(x)\dfrac{\partial^2}{\partial x^2}\dfrac{1}{((x+\nu)^2
+n^2)^{\,s}}\,dx\,.
\end{split}
\eeq{pds2} By definition
$\sum\limits_{\substack{n_i=1\\i=1,\ldots,d-1}}^{\infty}
\dfrac{1}{n^{2s}} = P_{d-1}(s)$. Therefore \beq
P_d(s)=\dfrac{\alpha(s)}{2}\,P_{d-1}(s-\tfrac{1}{2})
-\dfrac{1}{2}\,P_{d-1}(s) +R_d(s) \eeq{pds3} where $R_d(s)$ is the
remainder defined by \beq R_d(s) \equiv
\sum_{n_1,\ldots,n_{d-1}=1}^{\infty}
\dfrac{-1}{2}\sum_{\nu=0}^{\infty}\int_0^1
B_2(x)\dfrac{\partial^2}{\partial x^2}\dfrac{1}{((x+\nu)^2
+n^2)^{\,s}}\,dx\,. \eeq{rds} The remainder $R_d(s)$ is worked out
in appendix C and the result is \beq R_d(s)=
\sum_{n_1,\ldots,n_{d-1}=1}^{\infty}\sum_{\ell=1}^{\infty}\dfrac{2}{\sqrt{\pi}}\left(\dfrac{\pi\,\ell}{n}\right)^{\!s-1/2}
\!\!\!\Gamma(1-s)\,\sin(\pi\,s)\,K_{s-\!\!1/2}(2\,\pi\,\ell\,n)
\eeq{rds2} where $n$ is given by \reff{asquared}. We now evaluate
the term $2^d P_d(s)$ occurring in \reff{Zn} via  \reff{pds3} and
\reff{pkzm}: \beq
\begin{split}
2^d\,P_d(s)&= 2^{d-1}\,\alpha(s)\,P_{d-1}(s-\tfrac{1}{2}) - 2^{d-1}\,P_{d-1}(s) + 2^d\,R_d(s)\\&
=\sum_{m=1}^{d-1}(-1)^{d+m-1}\,\bino{d\!-\!1}{m}\left[\alpha(s)\,Z_m(s-\tfrac{1}{2}) - Z_m(s)\right] +2^d\,R_d(s)\,.
\end{split}
\eeq{2dpd} Substituting \reff{2dpd} into \reff{Zn} we obtain our
main equation: \beq
Z_d(s)=\sum_{m=1}^{d-1}(-1)^{d+m-1}\left[\alpha(s)\,\bino{d\!-\!1}{m}\,Z_m(s-\tfrac{1}{2})
+ \bino{d\!-\!1}{m\!-\!1}\,Z_m(s)\right] +2^d\,R_d(s) \eeq{Zdd}
where $R_d(s)$ is the remainder given by \reff{rds2}. Equation
\reff{Zdd} expresses $Z_d$ as sums over $Z_i$'s from $1$ to $d-1$
plus a remainder. We are now in a position to obtain expressions for
$Z_3,Z_5$ and $Z_7$ as products of one-dimensional sums plus a
remainder by using our main equation \reff{Zdd} together with the
analytical expressions for $Z_1,Z_2,Z_4,Z_6$ and $Z_8$ given in
\reff{zetas}. We begin with $Z_3(s)$. Applying equation \reff{Zdd}
yields \beq Z_3(s)=\alpha(s)\!\left[-2\,Z_1(s-\tfrac{1}{2})
+Z_2(s-\tfrac{1}{2})\right] -Z_1(s) +2\,Z_2(s) + 2^3\,R_3(s)\,.
\eeq{z3s} We now substitute the analytical expressions for $Z_1$ and
$Z_2$ given in \reff{zetas} and obtain our final expression for
$Z_3$: \beq
Z_3(s)=4\,\alpha(s)\,\zeta(s\!-\!\tfrac{1}{2})\beta(s\!-\!\tfrac{1}{2})
-4\,\alpha(s)\,\zeta(2s\!-\!1) + 8\,\zeta(s)\beta(s) -2\,\zeta(2s) +
8\,R_3(s)\,. \eeq{z3s2} This is a compact analytical result for the
important three-dimensional case. The only remainder is
 $8\,R_3(s)$ and the rest includes four analytical terms, each expressed in terms of simple one-dimensional sums and gamma functions.
Later we will see that the analytical part yields numerically the
correct Casimir energy to within $0.04\%$! We now evaluate $Z_5(s)$.
Using again the main equation \reff{Zdd} we obtain: \beq
\begin{split}
Z_5(s)&=\alpha(s)\left[-4\,Z_1(s\!-\!\tfrac{1}{2}) + 6\,Z_2(s-\tfrac{1}{2}) -4\,Z_3(s-\tfrac{1}{2}) + Z_4(s-\tfrac{1}{2})\right] \\&\qquad\qquad\qquad - Z_1(s) + 4\,Z_2(s)- 6\,Z_3(s)+ 4\,Z_4(s) + 2^5\,R_5(s)\,.
\end{split}
\eeq{z5s}
Substituting the analytical expressions for $Z_1,Z_2$ and $Z_4$ given in \reff{zetas} and $Z_3$ from \reff{z3s2} into \reff{z5s} one obtains the final expression for $Z_5(s)$:
\beq
\begin{split}
Z_5(s)& = 10\,\zeta(2s) -32\,\zeta(s)\beta(s) +32\,\zeta(s)\,\zeta(s-1)\,(1-4^{1-s}) \\&
+ 8\,\alpha(s)\left[3\,\zeta(2s-1)-4\,\zeta(s\!-\!\tfrac{1}{2})\beta(s\!-\!\tfrac{1}{2}) + \zeta(s)\,\zeta(s\!-\!\tfrac{3}{2})(1-2^{3-2s})\right]\\&
-16\,\alpha(s)\,\alpha(s\!-\!\tfrac{1}{2})\left(\zeta(s-1)\beta(s-1)-\zeta(2s-2)\right) + RZ_5(s)
\end{split}
\eeq{z5s2} where the remainder $RZ_5(s)=
-32\,\alpha(s)\,R_3(s\!-\!\tfrac{1}{2}) -48\,R_3(s) +32\,R_5(s)$.
The expression for $Z_7(s)$ is: \beq
\begin{split}
Z_7(s)&=\alpha(s)\left[-6\,Z_1(s\!-\!\tfrac{1}{2}) + 15\,Z_2(s-\tfrac{1}{2})- 20\,Z_3(s-\tfrac{1}{2})\right.\\&
\left. + 15\,Z_4(s-\tfrac{1}{2}) - 6\,Z_5(s-\tfrac{1}{2})+ Z_6(s-\tfrac{1}{2})\right]
-Z_1(s) + 6\,Z_2(s)\\&- 15\,Z_3(s)+ 20\,Z_4(s) - 15\,Z_5(s) + 6\,Z_6(s) + 2^7\,R_7(s)
\end{split}
\eeq{z7s2} where $Z_1,Z_2,Z_4$ and $Z_6$
 are given by \reff{zetas}, $Z_3$
by \reff{z3s2} and $Z_5$ by \reff{z5s2}. It would be cumbersome to
write out the analytical terms for $Z_7$ as we did for $Z_3$ and
$Z_5$. For calculations, one simply evaluates the necessary $Z$'s
and substitutes them in \reff{z7s2}. This ends our results for the
odd-dimensional Epstein zeta functions. One could have continued and
obtained expressions for $Z_9(s)$ but this is no longer interesting
as the expressions become way too long. We now state and discuss the
numerical results for the Casimir energy.

\section{Numerical results and discussion}

Table 1 contains the numerical results for the Casimir energy
density for periodic ($\varepsilon_p$), Dirichlet($\varepsilon_D$)
and Neumann($\varepsilon_N$) for $q$ large dimensions and $d-q$
dimensions of equal length $L$. This is calculated using the
formulas in \reff{periodicfinal3} and the equation \reff{rjfinal}
for the remainder $R_j(q)$ ($v$ and $L$ are assumed to be unity). We
state the analytical and remainder contribution separately and
calculate their sum to obtain the Casimir energy density. For
dimensions up to $d=5$, we include all values of $q$. For higher
dimensions up to $d=10$ we only state $q=0$. For numerical results
for the case where one has arbitrary lengths the reader is referred
to \cite{Neto,Li,Maclay}. The formulas derived in appendix B are
actually very well suited for such a numerical study but length
limitations restrict us here.

Table 1 shows that the absolute value of the Casimir energy density
for the periodic case is the largest, followed by the Neumann and
Dirichlet. Note that the sign in the Dirichlet case alternates in
two fashions: for a given $q$, it alternates as the dimension $d$
changes and it also alternates as $q$ changes for a given $d$. The
Casimir energy densities agree with a few exceptions with results
obtained by computing the Epstein zeta function and quoted in the
table in \cite{Wolfram}. For periodic boundary conditions, results
for $d=p$ (corresponding to $q=0$ in our case) are close to our
values but do not fully agree. For $d=2$ the values agree but for
$d=3$ they obtain $-0.81$ while we obtain $-0.838$. For $d=4$, they
obtain $-0.85$ while we obtain $-0.932$ and for $d=5$ they obtain
$-0.95$ while we obtain $-1.022$. Which values are correct? Table 3
contains an independent determination of the Casimir energy density
for the case $q=0$ for periodic boundary conditions. The values in
Table 3 for $d=3,4$ and $5$ are $-0.837537, -0.932077$ and
$-1.02283$ respectively and these values are in agreement with our
results. Therefore, in the few places where our results differ from
\cite{Wolfram}, our numerical values can be considered correct. Some
numerical results are also quoted for Dirichlet boundary conditions
in \cite{Neto,Li} where Epstein zeta functions were also used. In
\cite{Neto}, the column $u=0$ corresponds to our $q=0$ and are in
agreement. In \cite{Li} where $D$ is the spacetime dimension i.e.
$D=d+1$, their first column corresponds to our $d-q=2$ results and
are in agreement.

In Table 2 the percentage of the Casimir energy which is a remainder
is quoted for the different boundary conditions as a function of the
dimension $d$ (for simplicity, we quote the hypercube case $q=0$ but
the same trend is followed by all $q$ values). Table 2 confirms the
predictions made in section 3. Moving down the table, as the
dimension increases, the percentage decreases for Dirichlet but
increases for Neumann and periodic as predicted in section 3. Moving
horizontally across the table the percentage is lowest for Dirichlet
and largest for periodic with Neumann in between, again as predicted
in section 3 (with the only exception being $d=2$ due to the limited
low-energy permutations in the periodic and Neumann case and the
fact that the Dirichlet starts off at a low energy unlike higher
dimensions).

Note how small is the percentage remainder. Only at the highest
dimensions is the percentage high and this mostly for the periodic
case. The percentage remainder is negligible for the Dirichlet case
and the analytical formulas are all we need. The Neumann case has a
very low remainder at low dimensions. At $d=4$ it has less than a
$1\%$ remainder so that the analytical formulas are simply excellent
at lower dimensions. Even the periodic case at $d=3$ has only a
$1.6\%$ remainder but the remainder grows rapidly with dimension
compared to the other two cases.

Table 3 contains the Casimir energy for the periodic case at $q=0$
for values of $d$ ranging from $2$ to $8$ calculated via the
expressions for the homogeneous Epstein zeta function $Z_d(s)$
(again $v$ and $L$ are assumed to be unity). Our aim here was not to
make a complete table of Casimir values using the Epstein zeta
function. This has already been successfully done in \cite{Wolfram}.
The goal was mainly to calculate the analytical and remainder terms
for the homogeneous Epstein zeta function in $3,5$ and $7$
dimensions. For even dimensions, the expressions are calculated via
\reff{zetas} where there is no remainder. For the odd cases of $3,5$
and $7$ dimensions they are calculated via our derived expressions
\reff{z3s2},\reff{z5s2} and \reff{z7s2} and \reff{rds2} for the
remainder $R_d(s)$.  Note how close are the derived Epstein zeta
analytical results to the actual Casimir energy and hence the small
remainder percentage wise. The analytical expressions
\reff{z3s2},\reff{z5s2} and \reff{z7s2} we derived for the Epstein
zeta are limited to a few dimensions but are exceptionally accurate.
As already stated, for the realistic three-dimensional case, the
remainder is only a remarkable $0.04\%$ of the Casimir energy. As
one can see, the remainder for these few cases is smaller than the
remainder from our cut-off technique. The reason is due to the fact
that the odd-dimensional cases are derived from the even ones which
contain no remainder.

\begin{table}
\begin{center}
\caption{analytical and remainder contributions to Casimir energy
density for periodic, Dirichlet and Neumann boundary conditions}
\includegraphics[scale=0.65]{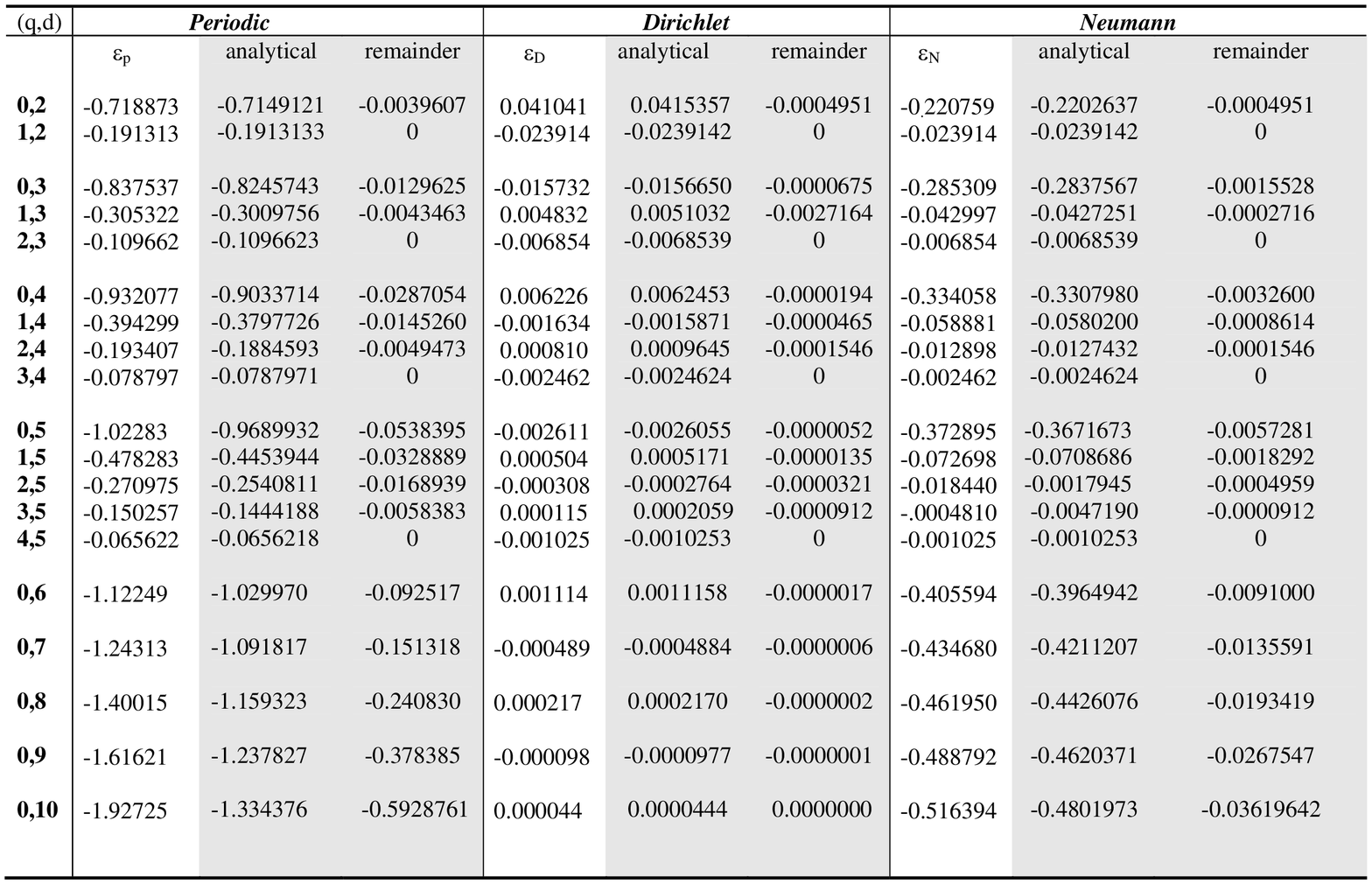}
\end{center}
\end{table}
\begin{table}
\begin{center}
\caption{percentage of Casimir energy which is remainder (case q=0)}
\includegraphics[scale=0.70]{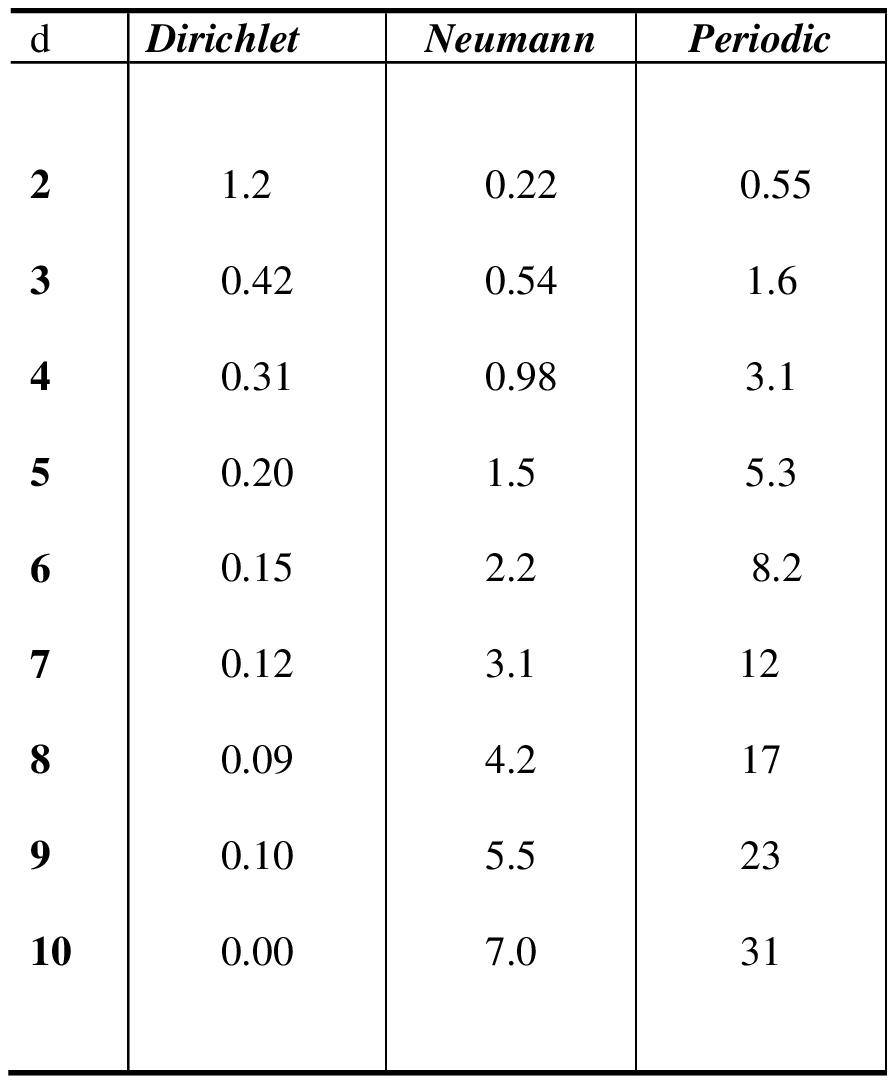}
\end{center}
\end{table}
\begin{table}
\begin{center}
\caption{Epstein-zeta function and comparison of remainder with
cut-off}
\includegraphics[scale=0.70]{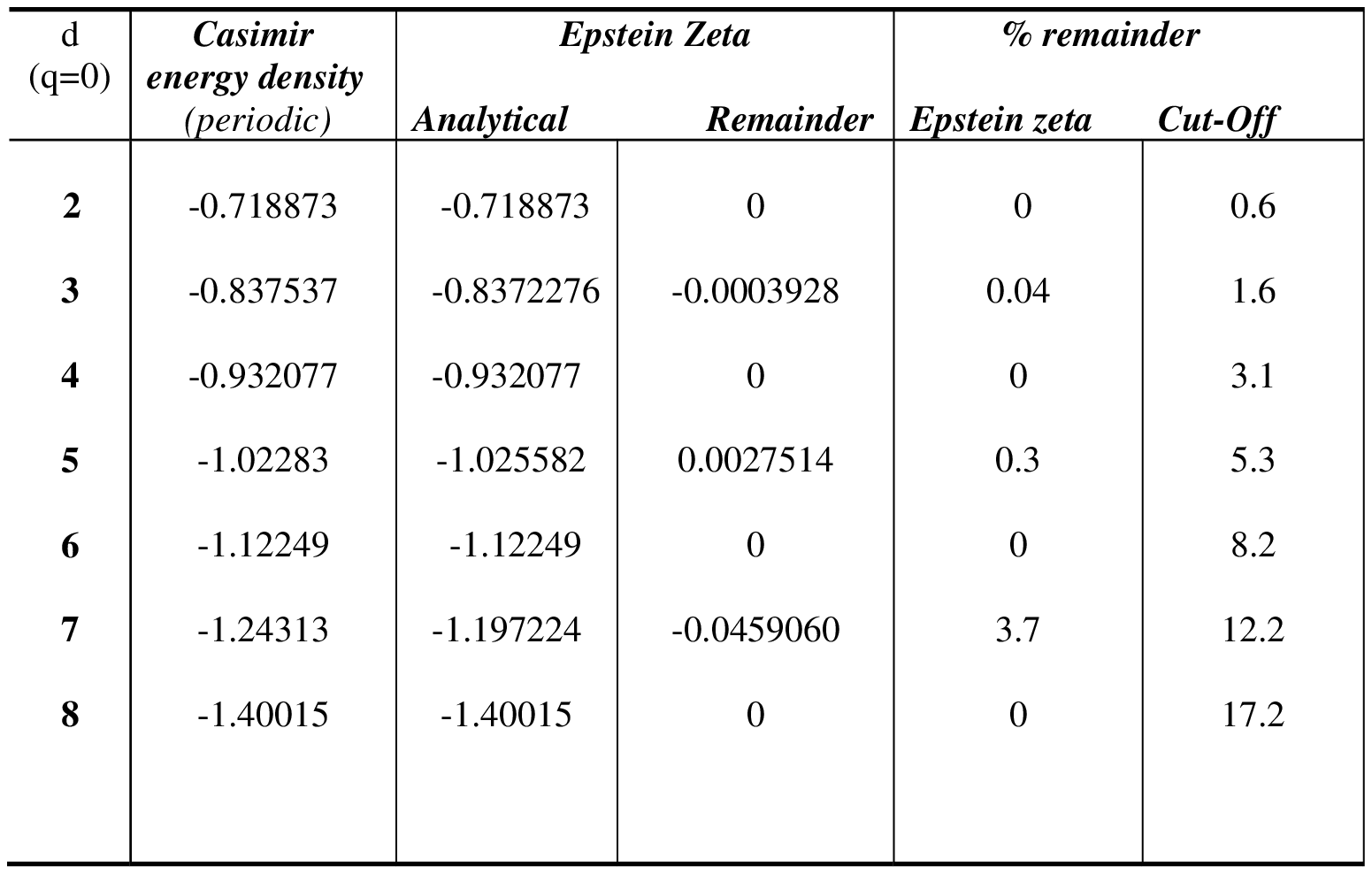}
\end{center}
\end{table}

\section*{\center{Acknowledgments}} I wish to thank the Natural
Sciences and Engineering Council of Canada (NSERC) and the Senate
Research Committee of Bishop's University for their financial
support of this project.

\begin{appendix}
\section{Remainder term $R_j(q)$}

In this appendix we evaluate the remainder term $R_j(q)$ defined by
\beq R_j(q) \equiv \lim_{\lambda\to 0}\, \partial_{\lambda}
\,R_j(q,\lambda) \eeq{rrj} where $R_j(q,\lambda)$ is given by
\reff{Rjlambda5} i.e. \beq
\begin{split}
R_j(q,\lambda) =&\sum_{m=1}^{j-q} \, \sum_{n=1}^{\infty}(-1)^m
\bino{j-q}{m}\,2^{j-m+1}\int_{0}^{\infty}\,\int_0^1\,\prod_{i=1}^{m}
\sum_{\nu_i=0}^{\infty} B_2(x_i) \,\dfrac{\partial^2}{\partial x_i}\\
&\expo{-\lambda\,\sqrt{n^2+ (x_1+\nu_1)^2 + \cdots + (x_m+\nu_m)^2 + y_1^2 + \cdots + y_{j-m}^2}}\, dx_1\ldots
dx_m\,dy_1\ldots dy_{j-m} \,.
\end{split}
\eeq{Rjlamb} There are $m$ integrals from $0$ to $1$ over the $x$'s
and $j-m$ integrals from $0$ to $\infty$ over the $y$'s. Our goal is
to simplify \reff{Rjlamb} as much as possible and put it in a
compact form useful for computations. In the end, the result is that
\reff{Rjlamb} can conveniently be reduced to sums over Bessel
functions. The first step is to convert the multiple integrals over
the $y$'s to a single integral by using spherical coordinates: \beq
r^2=y_1^2 + \cdots + y_{j-m}^2\,;\,dy_1\ldots dy_{j-m} =
2^{m-j+1}\,\dfrac{\pi^{\frac{j-m}{2}}}{\Gamma(\frac{j-m}{2})}\,r^{j-m-1}\,dr\,.
\eeq{spherical} $R_j(q,\lambda)$ is then reduced to \beq
\begin{split}
R_j(q,\lambda) =&\sum_{m=1}^{j-q} \,\sum_{n=1}^{\infty}(-1)^m \,
4\,\bino{j-q}{m}\,\dfrac{\pi^{\frac{j-m}{2}}}{\Gamma(\frac{j-m}{2})}
\int_{0}^{\infty}\,\int_0^1\,\prod_{i=1}^{m} \sum_{\nu_i=0}^{\infty}
B_2(x_i) \,\dfrac{\partial^2}{\partial x_i}\\&
\expo{-\lambda\,\sqrt{n^2+ (x_1+\nu_1)^2 + \cdots + (x_m+\nu_m)^2 +
r^2}}\, dx_1\ldots dx_m\,r^{j-m-1}\,dr\,.
\end{split}
\eeq{Rjbda3} We now turn to the x-integrals from $0$ to $1$. Note that $x+\nu$ is continuous and runs from $0$
to $\infty$. It is therefore convenient to drop the sum over $\nu$, replace $x+\nu$ by $x$ and integrate from
$0$ to $\infty$ instead of $0$ to $1$. This is valid as long as the Bernoulli function $B_2(x)$ is replaced by
$B_2(x-[x])$ where $[x]$ is the greatest integer less than or equal to $x$. This ensures that the Bernoulli
function is periodic with period $1$ while $x$ runs to infinity. Moreover, $B_2(0) = B_2(1)$ so that
$B_2(x-[x])$ is not only periodic but continuous. A fourier expansion of $B_2(x)=x^2-x+1/6$ can readily be
obtained and is given by
 \beq x^2-x+1/6= \sum_{\ell=1}^{\infty} \dfrac{\cos(2\,\pi\,\ell\,x)}{\ell^2\,\pi^2}\,.
\eeq{fourier} The right hand side of \reff{fourier} is a continuous
periodic function valid for all $x$. It is equal to the left hand
side only for $0\le x \le 1$ but equal to $B_2(x-[x])$ over the
entire region of integration $0\le x < \infty$. We can therefore
make the following replacement: \beq \sum_{\nu_i=0}^{\infty}
\int_{0}^1\,B_2(x_i) \,\dfrac{\partial^2\,f(x_i+\nu_i)}{\partial
x_i} dx_i \to
\sum_{\ell_i=1}^{\infty}\dfrac{1}{\ell_i^2\,\pi^2}\,\int_{0}^{\infty}
\cos(2\,\pi\,\ell_i\,x_i) \dfrac{\partial^2\,f(x_i)}{\partial x_i}
dx_i \eeq{replace} where $f(x_i)$ is the exponential function in
\reff{Rjbda3} with $\nu$ omitted i.e. \beq f(x_i)
=\expo{-\lambda\,\sqrt{n^2+ x_1^2 + \cdots + x_i^2 +\cdots+ x_m^2+
r^2}}\,. \eeq{f} The function $f$ has the following properties: \beq
\lim_{x_i\to 0} \dfrac{\partial\,f(x_i)}{\partial x_i} =0
\,;\,\lim_{x_i\to \infty} \dfrac{\partial\,f(x_i)}{\partial x_i} =0
\,;\,  \lim_{x_i\to \infty} f(x_i) =0\,. \eeq{properties} After
integrating by parts twice and using the above properties of $f$,
\reff{replace} reduces to \beq
\begin{split}&\int_{0}^{\infty} \cos(2\,\pi\,\ell_i\,x_i)
\dfrac{\partial^2\,f(x_i)}{\partial x_i} dx_i = -4\,\pi^2\ell_i^2
\int_{0}^{\infty} \cos(2\,\pi\,\ell_i\,x_i)\,f(x_i)\\&
\word{and}\sum_{\nu=0}^{\infty} \int_{0}^1\,B_2(x)
\,\dfrac{\partial^2\,f(x+\nu)}{\partial x} \,dx \to
-4\,\sum_{\ell=1}^{\infty}\,\int_{0}^{\infty} \cos(2\,\pi\,\ell\,x)
f(x) \,dx\,. \end{split}\eeq{replace3A} Substituting
\reff{replace3A} into equation \reff{Rjbda3} yields \beq
\begin{split}
R_j(q,\lambda) =&\sum_{m=1}^{j-q} \sum_{n=1}^{\infty}
\sum_{\ell_{1,\ldots,m}=1}^{\infty}\,4^{m+1}\,\bino{j-q}{m}
\dfrac{\pi^{\frac{j-m}{2}}}{\Gamma(\frac{j-m}{2})}\int_{0}^{\infty}\,\prod_{i=1}^{m} \cos(2\,\pi\,\ell_i\,x_i)\\
&\expo{-\lambda\,\sqrt{n^2+ x_1^2 + \cdots + x_m^2 + r^2}}\, dx_1\ldots dx_m\,r^{j-m-1}\,dr \,.
\end{split}
\eeq{Rj} We can reduce the above expression \reff{Rj} to sums over
the modified Bessel function $K_{\frac{j+1}{2}}$ by applying
sequentially the following set of three integrals \cite{Gradshteyn}:
\begin{flushleft}
I. $\int_0^{\infty} \cos(\gamma
\,x)\,\expo{-\lambda\,\sqrt{b^2+x^2}}\,dx =
\dfrac{\lambda\,b}{\sqrt{\lambda^2+\gamma^2}}
\,K_{-1}(b\,\sqrt{\lambda^2+\gamma^2})$\\
II. $\int_0^{\infty}(x^2+b^2)^{\mp\tfrac{1}{2}\nu}\,K_{\nu}(a\sqrt{x^2+b^2})\,\cos(c\,x)$\\
  $\qquad\qquad\qquad\qquad\qquad=(\tfrac{\pi}{2})^{1/2}\,\,a^{\mp\nu}\,b^{\tfrac{1}{2}\mp\nu}\,
  (a^2+c^2)^{\pm\tfrac{1}{2}\nu-\tfrac{1}{4}}\,\,K_{\pm\nu-\tfrac{1}{2}}(b\sqrt{a^2+c^2}).$\\
III.  $\int_0^{\infty}
K_{\nu}(\alpha\,\sqrt{z^2+x^2})\,\dfrac{x^{2\mu+1}}{(z^2+x^2)^{\nu/2}}\,dx
=
  \dfrac{2^{\mu}\,\Gamma(\mu +1)}{\alpha^{\mu+1}\,z^{\nu-\mu-1}}\,K_{\nu-\mu-1}(\alpha\,z)\,.$
\end{flushleft}
Integral I is applied once and converts the exponential and one
cosine into the modified Bessel function $K_{-1}$ i.e. \beq
\int_0^{\infty}
\cos(2\,\pi\,\ell_1\,x_1)\,\expo{-\lambda\,\sqrt{b^2+x_1^2}}\,dx_1 =
\dfrac{\lambda\,b}{\sqrt{\lambda^2+ 4\,\pi^2\,\ell_1^2}}
\,K_{-1}\big(\,b\,\sqrt{\lambda^2+ 4\,\pi^2\,\ell_1^2}\,\big)
\eeq{k1} where $b\equiv\sqrt{x_2^2+\cdots+x_m^2+n^2+r^2}$. We now
make repeated application of integral II for the remaining $x$'s
that appear in the definition of $b$. The subscript of the Bessel
function is therefore decreased by $1/2$ each time. Since there are
$m-1$ $x$-integrals to perform, and we start with $K_{-1}$, this
yields the Bessel function $K_{\tfrac{-m-1}{2}}$ i.e.
\beq\begin{split} &\int_{0}^{\infty}
\dfrac{\lambda}{\sqrt{\lambda^2+ 4\,\pi^2\,\ell_1^2}}\,
\prod_{i=2}^{m}\,\cos(2\,\pi\,\ell_i\,x_i)\,\,b\,\,\,K_{-1}\big(\,b\,\sqrt{\lambda^2+
4\,\pi^2\,\ell_1^2}\,\,\big)\,dx_2\ldots dx_m\\&
=\dfrac{\lambda\,(n^2+r^2)^{\tfrac{m+1}{4}}} {\pi\, 2^m\!
\left(\tfrac{\lambda^2}{4\pi^2}+\ell_1^2+\cdots+\ell_m^2\right)^{\!\!\!\tfrac{m+1}{4}}}
\,\,K_{\tfrac{-m-1}{2}}\big(\,2\pi
\sqrt{n^2+r^2}\sqrt{\tfrac{\lambda^2}{4\pi^2}+\ell_1^2+\cdots+\ell_m^2}\,\big)
\end{split}
\eeq{mint} We now apply integral III to perform the integration over
$r$  i.e. \beq
\begin{split}&\int_0^{\infty}\!\dfrac{\lambda\,\pi^{-1}\,2^{-m}\,(n^2+r^2)^{\tfrac{m+1}{4}}}{\left(\tfrac{\lambda^2}{4\pi^2}+\ell_1^2+\cdots+\ell_m^2\right)^{\!\!\!\tfrac{m+1}{4}}}
 \,\,\,K_{\tfrac{-m-1}{2}}\big(\,2\pi
\sqrt{n^2+r^2}\sqrt{\tfrac{\lambda^2}{4\pi^2}+\ell_1^2+\cdots+\ell_m^2}\,\,\big)\,r^{j-m-1}\,dr\\&\quad\quad\quad=
\dfrac{\lambda}{\pi\,2^{\,m+1}}\,\,\dfrac{\Gamma\big(\tfrac{j-m}{2}\big)}{\pi^{\tfrac{j-m}{2}}}\,\,\,
\dfrac{n^{\tfrac{j+1}{2}}\,K_{\tfrac{j+1}{2}}\big(\,2\pi\,n
\sqrt{\tfrac{\lambda^2}{4\pi^2}+\ell_1^2+\cdots+\ell_m^2}\,\,\big)}{\left(\tfrac{\lambda^2}{4\pi^2}+\ell_1^2+\cdots+\ell_m^2\right)^{\!\!\!\tfrac{j+1}{4}}}
\end{split}\eeq{kj7}

The integrals over $x$ and $r$ appearing in \reff{Rj} can now be
replaced by \reff{kj7} yielding:
 \beq R_j(q,\lambda) =
\dfrac{\lambda}{\pi} \sum_{m=1}^{j-q} 2^{\,m+1}\,\binom{j-q}{m}\,
\sum_{n=1}^{\infty}\sum_{\ell_{1,\ldots,m}=1}^{\infty}
\dfrac{n^{\frac{j+1}{2}}\,K_{\frac{j+1}{2}}(2\,\pi\,n\,\sqrt{\frac{\lambda^2}{4\pi^2}+\ell_1^2
+\cdots+\ell_m^2})}{(\frac{\lambda^2}{4\pi^2}+\ell_1^2 +\cdots+
\ell_m^2)^{\tfrac{j+1}{4}}}\,. \eeq{rjfinal2} Finally, by taking the
derivative of $R_j(q,\lambda)$ with respect to $\lambda$ and taking
the limit as $\lambda\to 0$ yields our desired final result for the
remainder $R_j(q)$: \beq
\begin{split}
 R_j(q) &\equiv \lim_{\lambda\to 0}\, \partial_{\lambda} \,R_j(q,\lambda)\\&
 = \dfrac{1}{\pi} \sum_{m=1}^{j-q}
2^{\,m+1}\,\binom{j-q}{m}\,
\sum_{n=1}^{\infty}\sum_{\ell_{1,\ldots,m}=1}^{\infty}
\dfrac{n^{\frac{j+1}{2}}\,K_{\frac{j+1}{2}}(2\,\pi\,n\,\sqrt{\ell_1^2
+\cdots+\ell_m^2})}{(\ell_1^2 +\cdots+
\ell_m^2)^{\tfrac{j+1}{4}}}\,.
\end{split}
\eeq{Rjfinitio} Our final expression \reff{Rjfinitio} for $R_j(q)$
is excellent for numerical calculations because it converges very
quickly (exponentially fast). The sums to infinity are formalities
as one can reach an accuracy of 8 to 10 digits by summing fewer than
9 numbers in each sum for $j$ up to $10$.

\section{Casimir energy in rectangular cavities with arbitrary lengths}

One can generalize the multidimensional cut-off method used in
section 2 to obtain Casimir energy formulas for arbitrary lengths in
a $d$-dimensional rectangular cavity. Our analysis will naturally be
brief since it follows closely that of section 2 and many results
from that section can be applied here. The best way to read this
appendix is therefore to have section 2 and appendix A in hand for
immediate reference.

The quantized frequencies $\omega$ for periodic (p), Neumann (N) and
Dirichlet (D) conditions are now given by:
 \beq
\begin{split}
&\omega_p = 2\pi\,v \,(\tfrac{n_1^2}{L_1^2} + \cdots +
\tfrac{n_{d}^2}{L_d^2})^{1/2} \\& \omega_{N,D} = \pi\,v
(\tfrac{n_1^2}{L_1^2} + \cdots + \tfrac{n_{d}^2}{L_d^2})^{1/2}
\end{split}
\eeq{omegaB} where the lengths range from $L_1$ to $L_d$. The
regularized vacuum energy for periodic boundary conditions is then
given by a similar form to \reff{def2} i.e. \beq\begin{split}
E^{\,reg}_p(\lambda) &= -\pi\,v \,\partial_{\lambda}\!
\sum_{\substack{n_i=-\infty\\i=1,\ldots, d}}^{\infty}
\expo{-\lambda\,\sqrt{\tfrac{n_1^2}{L_1^2} \,+\, \cdots\, +\,
\tfrac{n_d^2}{L_d^2}}}=-\pi\,v \,\partial_{\lambda}\,\Big(1\! +\!
\sumprime_{n_1=-\infty}^{\infty}\!\!\expo{-\lambda\,\sqrt{\tfrac{n_1^2}{L_1^2}}}\\&\quad
\!+\!\sumprime_{n_2=-\infty}^{\infty}\sum_{n_1=-\infty}^{\infty}\!\!\expo{-\lambda\,\sqrt{\tfrac{n_1^2}{L_1^2}
\,+\, \tfrac{n_2^2}{L_2^2}}} \,+\, \cdots \,+\,
\sumprime_{n_d=-\infty}^{\infty}\sum_{\substack{n_i=-\infty\\i=1,\ldots,
d-1}}^{\infty}\!\!\!\expo{-\lambda\,\sqrt{\tfrac{n_1^2}{L_1^2}\, +\,
\cdots \,+\, \tfrac{n_d^2}{L_d^2}}}\,\Big)\\&= -\pi\,v
\,\sum_{j=0}^{d-1}\partial_{\lambda}\,\Lambda_j(\lambda)
\end{split}\eeq{start2B}
where \beq
\Lambda_j(\lambda)\equiv\sumprime_{n=-\infty}^{\infty}\,\sum_{\substack{n_i=-\infty\\i=1,\ldots,
j}}^{\infty}\!\!\!\expo{-\lambda\,\sqrt{\tfrac{n^2}{L_{j+1}^2}\,+\,\tfrac{n_1^2}{L_1^2}
\,+\, \cdots + \tfrac{n_j^2}{L_j^2}}}\,. \eeq{lambdajB} As in
\reff{firstsum}, we obtain via the Euler-Maclaurin formula that \beq
\sum_{n_i=-\infty}^{\infty} f(n_i) =\int_{-\infty}^{\infty} f(x)\,dx
- R \,.\eeq{firstB} $R$ is given by expression \reff{replace3A}
obtained in appendix A: \beq
\begin{split}R&=\sum_{\nu=0}^{\infty} \int_{0}^1\,B_2(x)
\,\dfrac{\partial^2\,f(x+\nu)}{\partial x} \,dx \\&=
-4\,\sum_{\ell=1}^{\infty}\,\int_{0}^{\infty} \cos(2\,\pi\,\ell\,x)
f(x) \,dx = -2\,\sumprime_{\ell=-\infty}^{\infty}\,\int_{0}^{\infty}
\cos(2\,\pi\,\ell\,x) f(x) \,dx\end{split}\eeq{replace3B} where we
used $f(x)=f(-x)$ for the function we are considering. Then
\reff{firstB} reduces to \beq \sum_{n_i=-\infty}^{\infty} f(n_i) =
2\,\sum_{\ell=-\infty}^{\infty}\int_{0}^{\infty}
\cos(2\,\pi\,\ell\,x)  f(x)\,dx \eeq{twoB} where $\ell=0$ is now
included. Therefore the $j$-dimensional sum appearing in
\reff{lambdajB} for $\Lambda_j(\lambda)$  can be obtained by
repeated application of \reff{twoB}. What appears in the regularized
energy \reff{start2B} is the derivative
$\partial_{\lambda}\,\Lambda_j(\lambda)$: \beq\begin{split}
&\partial_{\lambda}\,\Lambda_j(\lambda)=
\partial_{\lambda}\sumprime_{n=-\infty}^{\infty}\sum_{\substack{n_i=-\infty\\i=1,\ldots,
j}}^{\infty}\!\!\!\expo{-\lambda\,\sqrt{\tfrac{n^2}{L_{j+1}^2}\,+\,\tfrac{n_1^2}{L_1^2}
\,+\, \cdots +
\tfrac{n_j^2}{L_j^2}}}\\&=\partial_{\lambda}\!\!\sumprime_{n=-\infty}^{\infty}\!\!2^{\,j}\!\!\!\sum_{\substack{l_i=-\infty\\i=1,\ldots,
j}}^{\infty} \int_{0}^{\infty}\!\!\!
\cos(2\,\pi\,\ell_1\,x_1)\ldots\cos(2\,\pi\,\ell_j\,x_j)\,\expo{-\lambda\,\sqrt{\tfrac{n^2}{L_{j+1}^2}\,+\,\tfrac{x_1^2}{L_1^2}
\,+\, \cdots + \tfrac{x_j^2}{L_j^2}}}\,dx_1\ldots
dx_j\\&=\dfrac{L_1\ldots L_j}{(L
_{j+1})^{j+1}}\,\Big(\,2^{\,j+1}\,\partial_{\lambda}\!\!\sum_{n=1}^{\infty}\int_{0}^{\infty}\,\expo{-\lambda\,\sqrt{n^2+
x_1^2 \,+\, \cdots + x_j^2}}\,dx_1\ldots dx_j +
\partial_{\lambda}R_j(\lambda)\,\Big)\end{split}\eeq{jsum} where the sum over all
$\ell$'s was divided into two cases leading to the two terms in the
brackets of \reff{jsum}. The first term occurs when all $\ell$'s are
equal to zero. The second term is for all other $\ell$'s and
corresponds to the remainder :\beq\begin{split}
&\partial_{\lambda}\,R_j(\lambda)\equiv\\&
2^{\,j+1}\,\partial_{\lambda}\!\!\sum_{n=1}^{\infty}\sumprime_{\substack{l_i=-\infty\\i=1,\ldots,
j}}^{\infty}\!\! \int_{0}^{\infty}\!\!\!\!\!
\cos(2\pi\ell_1\tfrac{L_1}{L_{j+1}}\,x_1)\ldots\cos(2\pi\ell_j
\tfrac{L_j}{L_{j+1}}\,x_j)\,\expo{-\lambda\,\sqrt{n^2+ x_1^2 +
\cdots + x_j^2}}dx_1\ldots dx_j
\end{split}\eeq{RjlamB} where the prime over the multiple sum excludes only the case when all $\ell$'s are equal to
zero. The multiple integral over $j$ cosines can be obtained
directly from \reff{mint} in appendix A by the following
substitutions: $m \to j$, $\ell_i\to\ell_i\,L_i/L_{j+1}$ and
$n^2+r^2\to n^2$ i.e. \beq\begin{split}
&\partial_{\lambda}\,R_j(\lambda)=
\\&\partial_{\lambda}\!\sum_{n=1}^{\infty}\sumprime_{\substack{l_i=-\infty\\i=1,\ldots,
j}}^{\infty}\dfrac{2\,\lambda\,(n\,L_{j+1})^{\tfrac{j+1}{2}}K_{\tfrac{j+1}{2}}
\big(\tfrac{2\pi\,n}{L_{j+1}}\sqrt{\tfrac{(\lambda\,L_{j+1})^2}{4\pi^2}+(\ell_1\,L_1)^2+\cdots+(\ell_j^2\,L_j)^2}\,\big)}
{\pi\,
\left(\tfrac{(\lambda\,L_{j+1})^2}{4\pi^2}+(\ell_1\,L_1)^2+\cdots+(\ell_j\,L_j)^2\right)^{\!\!\!\tfrac{j+1}{4}}}
\end{split} \eeq{lmcs}
 The Casimir energy is proportional to the finite part
of \reff{jsum} as $\lambda \to 0$. The first term in brackets in
\reff{jsum} is identical to the derivative of the first term in
$\Lambda_j(q,\lambda)$ given by \reff{Lambda7}. Therefore the result
\reff{lambdafine} from section 2 is directly applicable i.e. \beq
\lim_{\lambda\to
0}\partial_{\lambda}\,\Lambda^{finite}_j(\lambda)=\dfrac{L_1\ldots
L_j}{(L _{j+1})^{j+1}}\Big(
\Gamma(\tfrac{j+2}{2})\,\pi^{\frac{-j-4}{2}}\, \zeta(j+2) + R_j
\Big) \eeq{lambdafineB} where the remainder term is given by
\beq\begin{split} R_j &\equiv \lim_{\lambda\to
0}\,\partial_{\lambda}\,R_j(\lambda)\\&=\sum_{n=1}^{\infty}\,\sumprime_{\substack{l_i=-\infty\\i=1,\ldots,
j}}^{\infty}\dfrac{2\,(n\,L_{j+1})^{\tfrac{j+1}{2}}} {\pi\,
\left[(\ell_1\,L_1)^2+\cdots+(\ell_j\,L_j)^2\right]^{\tfrac{j+1}{4}}}\,K_{\tfrac{j+1}{2}}
\big(\,\tfrac{2\pi\,n}{L_{j+1}}\sqrt{(\ell_1\,L_1)^2+\cdots+(\ell_j\,L_j)^2}\,\,\,\big)\,.
\end{split}
\eeq{rjBB} Our final Casimir energy expression for periodic boundary
conditions is then given by \beq
\begin{split}E_{p_{ _{L_1\ldots L_d}}}(d) &= -\pi\, v\sum_{j=0}^{d-1}\lim_{\lambda\to
0}\partial_{\lambda}\,\Lambda^{finite}_j(\lambda)\\&=-\pi\,
v\sum_{j=0}^{d-1}\dfrac{L_1\ldots L_j}{(L _{j+1})^{j+1}}\Big(
\Gamma(\tfrac{j+2}{2})\,\pi^{\frac{-j-4}{2}}\, \zeta(j+2) + R_j
\Big)\\&=v\,\Big(\,\dfrac{-\pi}{6\,L_1}-\dfrac{L_1}{L_2^2}\dfrac{\zeta(3)}{2\,\pi}-\dfrac{L_1\,L_2}{L_3^3}\dfrac{\pi^2}{90}+\cdots
-R_1\,\dfrac{\pi\,L_1}{L_2^2} -R_2\,\dfrac{\pi\,L_1\,L_2}{L_3^3}
+\cdots\Big)
\end{split}\eeq{epfinal} where the remainder $R_j$ is
given by \reff{rjBB} (note that $R_j$ is zero when $j=0$). Equation
\reff{epfinal} is a highly compact way to express the Casimir energy
for arbitrary lengths. As in section 2 it is split into two terms:
an analytical part and a remainder. The same physical interpretation
follows: the analytical part is a sum of parallel plate terms.
Equation \reff{epfinal} is valid for any lengths and we know the
result should be invariant under a permutation of the lengths.
However, the two terms separately are not invariant, only their sum.
We naturally want to label the lengths such that the remainder term
lives up to its name. This can be accomplished if the largest length
is labeled $L_1$, the next largest length $L_2$, i.e. $L_1\ge L_2\ge
L_3...$. Then the Bessel function decreases exponentially fast and
the remainder is small. If $q$ dimensions are large and $d-q$
dimensions have equal length $L$, Eq.\reff{epfinal} for $E_p(d)$ and
Eq.\reff{rjBB} for the remainder $R_j$ reduce to the results of
section 2 i.e. $E_p(q,d)$ given by \reff{periodicfinal2} and
$R_j(q,d)$ given by \reff{rjfinal} respectively.

The Neumann (N) and Dirichlet (D) cases can be obtained via simple
permutations of the periodic case. The operator relations for
Neumann and Dirichlet are $\sum_{0}^{\infty} \to \tfrac{1}{2} \bigl(
\sum_{-\infty}^{\infty} + 1\bigr)$ and $\sum_{1}^{\infty} \to
\tfrac{1}{2} \bigl( \sum_{-\infty}^{\infty} -1\bigr)$ respectively.
Applying the operator $d$ times while keeping each sum distinct
because of different lengths and multiplying the final result by
$\tfrac{1}{2}$ yields the Neumann and Dirichlet energies \beq
E_{N,D} = \dfrac{1}{2^{d+1}}\sum_{m=1}^d \sum_{(k_1,\ldots,
k_m)}\,(\pm 1)^{d+m}\,E_{p_{_{\,k_1\ldots k_m}}}(m) \eeq{yoops}
where the (+) is for Neumann and the (-) for Dirichlet. The sum is
over all sets $(k_1,\ldots, k_m)$ with
$k_1\!<\!k_2\!<\!\cdots\!<k_m$ (the $k$'s run from $1$ to $d$).
$E_{p_{_{k_1\ldots k_m}}}(m)$ is the periodic energy \reff{epfinal}
replacing $d$ by $m$ and $L_1$ by $L_{k_1}$, $L_2$ by $L_{k_2}$,
etc.

\section{Remainder term $R_d(s)$ for Epstein-zeta function}

We derive in this appendix a convenient form for the remainder
$R_d(s)$ in terms of sums of Bessel and gamma functions. We begin
with the expression for the remainder $R_d(s)$ given by \reff{rds}:
\beq R_d(s) \equiv \sum_{n_1,\ldots,n_{d-1}=1}^{\infty}
\dfrac{-1}{2}\sum_{\nu=0}^{\infty}\int_0^1
B_2(x)\dfrac{\partial^2}{\partial x^2}\dfrac{1}{((x+\nu)^2
+n^2)^{\,s}}\,dx\eeq{rdss} where \beq n^2\equiv n_1^2 + \cdots +
n_{d-1}^2\,. \eeq{nsquared} We now follow similar procedures as
those employed in appendix A for $R_j(q)$. To avoid being
repetitive, we skim through details already discussed in appendix A.

The term  $x+\nu$ is continuous and runs from $0$ to $\infty$. We
drop the sum over $\nu$, replace $x+\nu$ by $x$ and integrate from
$0$ to $\infty$ instead of $0$ to $1$. We replace $B_2(x)=x^2-x+1/6$
by its fourier expansion \reff{fourier} i.e.
 \beq x^2-x+1/6= \sum_{\ell=1}^{\infty} \dfrac{\cos(2\,\pi\,\ell\,x)}{\ell^2\,\pi^2}\,.
\eeq{fourier2} We can therefore make the following replacement in
\reff{rdss}: \beq \sum_{\nu=0}^{\infty} \int_{0}^1\,B_2(x)
\,\dfrac{\partial^2\,f(x+\nu)}{\partial x} dx \to
\sum_{\ell=1}^{\infty}\dfrac{1}{\ell^2\,\pi^2}\,\int_{0}^{\infty}
\cos(2\,\pi\,\ell\,x) \dfrac{\partial^2\,f(x)}{\partial x} dx
\eeq{replace2} where $f(x)$ is the function in \reff{rdss} with
$\nu$ omitted i.e. \beq f(x) = \dfrac{1}{(x^2+n^2)^s}\eeq{f2} The
function $f(x)$ has the following properties: \beq \lim_{x\to0}
\dfrac{\partial\,f(x)}{\partial x} =0 \,;\,\lim_{x\to \infty}
\dfrac{\partial\,f(x)}{\partial x} =0 \,;\,  \lim_{x\to \infty} f(x)
=0\,. \eeq{properties2} With the above properties of $f$,
\reff{replace2} reduces to the same expression \reff{replace3A}
obtained in appendix A: \beq \sum_{\nu=0}^{\infty}
\int_{0}^1\,B_2(x) \,\dfrac{\partial^2\,f(x+\nu)}{\partial x} \,dx
\to -4\,\sum_{\ell=1}^{\infty}\,\int_{0}^{\infty}
\cos(2\,\pi\,\ell\,x) f(x) \,dx \,.\eeq{replace3} After substituting
\reff{replace3} into \reff{rdss} we obtain $R_d(s)$ in the following
form: \beq R_d(s) =\sum_{n_1,\ldots,n_{d-1}=1}^{\infty}
\sum_{\ell=1}^{\infty}\,\int_{0}^{\infty} 2\,\cos(2\,\pi\,\ell\,x)
\dfrac{1}{(x^2 +n^2)^{\,s}}\,dx\,. \eeq{rdss2} The integral can be
expressed in terms of Bessel functions i.e.
 \beq
\int_{0}^{\infty} 2\,\cos(2\,\pi\,\ell\,x) \dfrac{1}{(x^2
+n^2)^{\,s}}\,dx =
\dfrac{2}{\sqrt{\pi}}\,\Gamma(1-s)\,\sin(\pi\,s)\,K_{s-1/2}(2\,\pi\,\ell\,n)\left(\dfrac{\pi\,\ell}{n}\right)^{s-1/2}\,.
\eeq{rdss3} Our final expression for $R_d(s)$ is then \beq R_d(s)
=\sum_{n_1,\ldots,n_{d-1}=1}^{\infty}
\sum_{\ell=1}^{\infty}\,\dfrac{2}{\sqrt{\pi}}\,\Gamma(1-s)\,\sin(\pi\,s)\,K_{s-1/2}(2\,\pi\,\ell\,n)\left(\dfrac{\pi\,\ell}{n}\right)^{s-1/2}
\eeq{rdsfinal} where $n\equiv \sqrt{n_1^2+\cdots+n_{d-1}^2}$ \,.
\end{appendix}


\begin{thebibliography}{99}
\bibitem{Casimir} H. G. Casimir, Proc. Kon. N. Akad. Wet. {\bf 51}, 793 (1948).
\bibitem{Spaarnay} M.J. Sparnaay, Physica {\bf 24}, 751 (1958).
\bibitem{Lamoreaux} S .K. Lamoreaux, Phys. Rev. Lett. {\bf 78}, 5 (1997).
\bibitem{Mohideen} U. Mohideen and A. Roy, Phys. Rev. Lett. {\bf 81}, 4549 (1998)
\bibitem{Klim} G. L. Klimchiskaya, A. Roy, U. Mohideen and V. M. Mostepanenko, Phys. Rev. A{\bf60}, 3487 (1999).
\bibitem{Roy1} A. Roy and U. Mohideen, Phys. Rev. Lett. {\bf 82}, 4380, (1999).
\bibitem{Roy2} A. Roy, C.-Y. Lin and U. Mohideen, Phys. Rev. D{\bf 60}, 111101(R) (1999).
\bibitem{Harris} B. W. Harris, F. Chen and U. Mohideen, Phys. Rev. A {\bf 62}, 052109 (2000).
\bibitem{Chen1} F. Chen, G. L. Klimchitskaya, U. Mohideen and V. M. Mostepanenko, Phys. Rev. A{\bf 69}, 022117 (2004).
\bibitem{Ederth} T. Ederth, Phys. Rev. A {\bf 62}, 062104 (2000).
\bibitem{Bressi} G. Bressi, G. Carugno, R. Onofrio and G. Ruoso, Phys. Rev. Lett. {\bf 88}, 041804 (2002).
\bibitem{Chen2} F. Chen, U. Mohideen, G. L. Klimchitskaya and V. M. Mostepanenko, Phys. Rev. Lett. {\bf 88}, 101801 (2002);
Phys. Rev. A {\bf 66}, 032113 (2002).
\bibitem{Decca} R. S. Decca, D. L\'{o}pez, E. Fischbach and D. E. Krause, Phys. Rev. Lett. {\bf 91}, 050402 (2003).
\bibitem{Dvali} N. Hamed, S. Dimopoulos and G. Dvali, Phys. Lett. B {\bf 429} 263 (1998).
\bibitem{Lopez} G. L. Klimchitskaya, R. S. Decca, E. Fischbach, D. E. Krause, D. L\'{o}pez, and V.M. Mostepanenko, Int. J. Mod. Phys. A{\bf 20}, 2205-2221 (2005).
\bibitem{Janina} J. Marciak-Kozlowska and M. Kozlowski, cond-mat/0506226.
\bibitem{Chan} H.B. Chan, V. A. Aksyuk, R. N. Kleiman, D. J. Bishop and F. Capasso, Science {\bf291}, 1941 (2001); Phys. Rev. Lett. {\bf 87}, 211801 (2001).
\bibitem{Roberts2} D.C. Roberts and Y. Pomeau, Phys. Rev. Lett. {\bf 95} 145303 (2005).
\bibitem{Roberts} D.C. Roberts and Y. Pomeau, cond-mat/0503757, 2005.
\bibitem{Pitaevskii} L. Pitaevskii and S. Stringari, Phys. Rev. Lett. {\bf 81}, 4541 (1999).
\bibitem{Stampur} D.M. Stampur Kurn {\it et. al}, Phys. Rev. Lett. {\bf 83}, 2876 (1999).
\bibitem{Greiner} M. Greiner {\it et. al.,} Nature {\bf 415}, 39 (2002).
\bibitem{Polchinski} J. Polchinski, {\it String Theory, Vol. I: An introduction to the Bosonic
String}, (Cambridge University Press, 1998).
\bibitem{Zee} A. Zee, {\it Quantum Field Theory in a Nutshell},
(Princeton University Press, 2003).
\bibitem{Zuber} C. Itzykson and J. B. Zuber, {\it Quantum Field
theory}, (McGraw-Hill, 1980).
\bibitem{GuthMIT} A. Guth, ``Relativistic Quantum Field Theory I: Spring 2003",
http://ocw.mit.edu/OcwWeb/Physics/8-323Relativistic-Quantum-Field-Theory-ISpring2003.
\bibitem{Vogels} J.M. Vogels, K. Xu and W. Ketterle, Phys. Rev. Lett. {\bf 89}, 020401 (2002); D.C. Roberts, T. Gasenzer and K. Burnett, J. Phys. B. {\bf 35}, L113-L118 (2002); H. Pu and P. Meystre, Phys. Rev. Lett. {\bf 85}, 3987 (2000); L.M. Duran {\it et. al.} Phys. Rev. Lett. {\bf 85}, 3991 (2000).
\bibitem{Kimball} K. A. Milton, {\it The Casimir Effect}, (World
Scientific, 2001).
\bibitem{Mostepanenko} V.M. Mostepanenko and N.N. Trunov, {\it The
Casimir effect and its applications}, (Oxford, 1997).
\bibitem{Kimball2} K. A. Mlton, J. Phys. A: Math. Gen., {\bf 37} 209
(2004).
\bibitem{Bordag} M. Bordag, U. Mohideen and V.M. Mostapanenko, Phys.Rept.{\bf 353} 1
(2001).
\bibitem{Barton} G. Barton in {\it Advances in Atomic and Molecular
Physics}, Suppl. 2, P.R. Berman, ed., (Academic Press, NY, 2004).
\bibitem{Jaeckel} M. Jaeckel and S. Reynaud, Rep. Prog. Physics {\bf
60} 863 (1997).
\bibitem{Visser} M. Visser, Class. Quant. Grav.{\bf 15} 1767 (1998).
\bibitem{Visser2} C. Barceló, S. Liberati and M. Visser, Class. Quantum Grav. {\bf 18}
1137 (2001).
\bibitem{Svaiter2} N.F. Svaiter and B.F. Svaiter, J. Math. Phys. {\bf 32}, 1 (1991).
\bibitem{Svaiter3} N.F. Svaiter and B.F. Svaiter, J. Phys. A: Math. Gen. {\bf 25}, 979 (1992).
\bibitem{Lukosz} W. Lukosz, Z. Phys. {\bf 262}, 327 (1973).
\bibitem{Trunov1} S.G. Mamayev and N.N. Trunov, Theor Math. Phys.(USA) 38
(1979).
\bibitem{Trunov2} V.M. Mostepanenko and N.N. Trunov, Sov. Phys.-- Usp.(USA) 31 (1988).
\bibitem{Beneventano} C.G. Beneventano and E.M. Santagelo, Int.J.Mod.Phys.A{\bf 11}, 2871 (1996).
\bibitem{Wolfram} J. Ambj{\o}rn and S. Wolfram, Ann. Phys. (N.Y.) {\bf 147}, 1 (1983).
\bibitem{Neto} F. Caruso, P. Neto, B.F. Svaiter and N.F. Svaiter, Phys. Rev. D {\bf 43},
1300 (1991).
\bibitem{Li} H. Cheng, X. Li, J. Li, and X. Zhai, Phys. Rev. D {\bf 56}, 2155 (1997).
\bibitem{Maclay} G. Maclay, Phys. Rev. A {\bf 61} 052110 (2000).
\bibitem{Elizalde2} E. Elizalde, Commun.Math.Phys.{\bf 198} 83
(1998).
\bibitem{Lee} T. D. Lee, K. Huang and C.N. Yang, Phys. Rev. {\bf 106}, 1135 (1957).
\bibitem{Bogo} N. Bogoliubov, J. Phys. (U.S.S.R.) {\bf 11}, 23 (1947).
\bibitem{Hardy} G.H. Hardy and E. M. Wright, {\it An Introduction to the Theory of Numbers,
5th ed.}, (Clarendon Press, 1979).
\bibitem{Goro} G. Shimura, Amer. J. Math. {\bf 124}, 1059 (2002).
\bibitem{Arfken} G. B. Arfken and H. J. Weber, {\it Mathematical Methods for Physicists,
4th edition}, (Academic Press, 1995).
\bibitem{Gradshteyn} I. S. Gradshteyn and I. M. Ryzhik, {\it Table of Integrals, Series and
Products, 6th edition},(Academic Press, 2000).
\bibitem{Ariel} A. Edery, J. Math. Phys. {\bf 44}, 599 (2003).
\bibitem{Ariel2} A. Edery, math-ph/0411056.
\bibitem{book1} K. Kirsten, {\it Spectral Functions in Mathematics
and Physics}, (Chapman \& Hall/CRC, 2001).
\bibitem{book2} E. Elizalde, {\it Ten Physical Applications of
Spectral Zeta Functions}, (Springer, 1995).
\bibitem{book3} E. Elizalde, S.D. Odintsov, A. Romeo, A.A. Bytsenko
and S. Zerbini, {\it Zeta Regularization Techniques with
Applications}, (World Scientific, 1994).
\bibitem{Kirsten1} G. Esposito, G. Fucci, A. Y. Kamenshchik and K.
Kirsten, Class. Quant. rav. {\bf 22} 957 (2005).
\bibitem{Elizalde3} G. Cognola, E. Elizalde and S. Zerbini, Phys.
Lett B{\bf 585} 155 (2004).
\bibitem{Elizalde4} E. Elizalde, S. Nojiri, S. Odintsov and S.
Ogushi, Phys. Rev. D{\bf 67} 063515 (2003).
\bibitem{Fulling} S. Fulling, J. Phys. A: Math. Gen. {\bf 36}, 6857 (2003).
\bibitem{Elizalde5} E. Elizade, J.Phys.A {\bf 34} 3025 (2001).
\bibitem{Kirsten2} G. esposito, P. Gilkey and K. Kirsten,
J.Phys.A {\bf 38} 2259 (2005).
\bibitem{Schakel} A. Schakel, J. Phys. Stud.{\bf 7} 140 (2003).
\bibitem{Xin} X. Li, X. Shi and J. Zhang, Phys. Rev. D{\bf 44} 560
(1991).
\bibitem{Ortenzi} G. Ortenzi and M. Speafico, J. Phys. A{\bf 37},
11499 (2004);
\bibitem{E1} E. Elizalde and A. Romeo, J. Math. Phys. {\bf 30}, 1133
(1989).
\bibitem{E2} E. Elizalde, J. Phys. A: Math. Gen. {\bf 22} 931(1989).
\bibitem{Elizalde6} E. Elizalde, J. Phys. A {\bf 22}, 931 (1989).
\bibitem{Cognola} G. Cognola, L. Vanzo and S. Zerbini, J. Math. Phys. {\bf 33}, 222 (1992).
\bibitem{Kirsten3} K. Kirsten, J.Phys. A: Math. Gen. {\bf 25}, 6297 (1992).
\bibitem{mathworld} E. Weisstein, {\it Sum of Squares Function}, MathWorld--A Wolfram Web\\
Resource. http://mathworld.wolfram.com/SumofSquaresFunction.html.
\end{thebibliography}
\end{document}